\documentclass[10pt,prd,twocolumn,floatfix,tightenlines,showpacs,showkeys,preprintnumbers,nofootinbib, longbibliography]{revtex4-1}

\usepackage{epsfig}
\usepackage{amsmath,bm,amsfonts}
\usepackage{hyperref}
\usepackage{multirow}
\usepackage{placeins}
\usepackage{color}

\newcommand{\w}{\omega}
\newcommand{\lam}{\lambda}

\definecolor{purple}{rgb}{0.63,0.13,0.94}
\definecolor{red}{rgb}{1.0,0.0,0.0}
\definecolor{green}{rgb}{0.0,1.0,0.0}
\definecolor{blue}{rgb}{0.0,0.0,1.0}
\definecolor{orange}{rgb}{0.8,0.6,0,0}
\definecolor{magenta}{rgb}{0.8,0.0,0.6}
\definecolor{white}{rgb}{1.0,1.0,1.0}
\definecolor{black}{rgb}{0.0,0.0,0.0}

\begin{document}

\title{Non-standard neutrino self-interactions in a supernova and fast flavor conversions}

\author{Amol Dighe}
\email{amol@theory.tifr.res.in}
\affiliation{Tata Institute of Fundamental Research, Homi Bhabha Road, Mumbai 400005, India}

\author{Manibrata Sen}
\email{manibrata@theory.tifr.res.in}
\affiliation{Tata Institute of Fundamental Research, Homi Bhabha Road, Mumbai 400005, India}

\date{\today}
\preprint{TIFR/TH/17-33}

\begin{abstract} 
We study the effects of non-standard self-interactions (NSSI) of neutrinos streaming out of a core-collapse supernova. We show that with NSSI, the standard linear stability analysis gives rise to linearly as well as exponentially growing solutions. For a two-box spectrum, we demonstrate analytically that flavor-preserving NSSI lead to a suppression of bipolar collective oscillations. In the intersecting four-beam model, we show that flavor-violating NSSI can lead to fast oscillations even when the angle between the neutrino and antineutrino beams is obtuse, which is forbidden in the Standard Model. This leads to the new possibility of fast oscillations in a two-beam system with opposing neutrino-antineutrino fluxes, even in the absence of any spatial inhomogeneities. Finally, we solve the full non-linear equations of motion in the four-beam model numerically, and explore the interplay of fast and slow flavor conversions in the long-time behavior, in the presence of NSSI. 
\end{abstract}


\maketitle

\section{Introduction}
\label{sec:intro}
Neutrinos exiting a core-collapse supernova (SN) can undergo rapid flavor conversions. Such flavor oscillations have been studied widely in the literature (for recent reviews, see \cite{Mirizzi:2015eza,Horiuchi:2017sku}). These conversions may play a crucial role during a SN explosion and in the formation of heavy elements during nucleosynthesis. Early works in this field focussed on the impact of  Mikheyev-Smirnov-Wolfenstein (MSW) effect on neutrino propagation \cite{Wolfenstein:1977ue,Mikheev:1986gs}, such that flavor conversions take place mainly in the resonance regions, where the matter potential is almost equal to the vacuum oscillation frequency \cite{Kuo:1989qe,Dighe:1999bi}. For typical SN matter profiles, this condition is satisfied at a radius of $r\sim\mathcal{O}(10^3)\,$km. 

However, in a dense gas of neutrinos/antineutrinos, the net neutrino-neutrino interactions are also significant, leading to non-linear collective effects \cite{Pantaleone:1992eq,Duan:2006an}.
Neutrinos undergo elastic forward scattering not only with the background matter, but also with the other neutrinos. As a result, they experience an effective potential due to the background neutrinos, and self-induced oscillations can take place \cite{Hannestad:2006nj,Horiuchi:2017sku,Duan:2010bg,Mirizzi:2015eza,Fogli:2007bk,EstebanPretel:2007ec,Chakraborty:2016yeg}. Such flavor conversions may occur earlier than the MSW conversions, at a radius of $r\sim\mathcal{O}(10^2)\,$km \cite{Duan:2006an,Hannestad:2006nj}, and give rise to new phenomena like synchronized oscillations, bipolar oscillations \cite{Hannestad:2006nj}, and spectral splits/swaps \cite{Raffelt:2007xt,Raffelt:2007cb,Fogli:2008pt,Dasgupta:2008cd,Dasgupta:2009mg}. While the non-linearity of the equations of motion (EoMs) makes the analytic understanding of long time behavior of collective oscillations intractable, it is possible to analytically study the onset of these oscillations using the linear stability analysis \cite{Banerjee:2011fj,Raffelt:2013rqa}.

Recently, it has been realised that ``fast flavor oscillations'', which can have frequencies a few orders of magnitude larger than the above bipolar collective oscillations (which we now refer to as ``slow oscillations''), may occur much closer to the SN core \cite{Sawyer:2005jk,Sawyer:2008zs,Sawyer:2015dsa}. These would take place if there exists a non-trivial zenith angle distribution of the different neutrino-antineutrino species \cite{Chakraborty:2016lct,Dasgupta:2016dbv}. Due to the differences in interaction cross-sections, the non-electron neutrinos would decouple earlier than the electron neutrinos and hence would indeed have a more forward peaked zenith angle distribution \cite{Tamborra:2017ubu}. Thus, these conversions can take place at a radius of $r\sim\mathcal{O}(10)\,$km \cite{Dasgupta:2016dbv}. Such conversions would lead to significant mixing of neutrino flavors as soon as they start free-streaming. Since these conversions take place deep inside a SN core, they can have an important impact on SN explosions and nucleosynthesis.

Fast flavor oscillations seem to require a ``crossing'' in the zenith angle distributions of the differences in the $\nu_e$ and $\bar{\nu}_e$ spectra \cite{Dasgupta:2016dbv,Sen:2017ogt}. This was further elucidated in \cite{Izaguirre:2016gsx,Capozzi:2017gqd} by using dispersion relations, which showed that a crossing in the electron lepton number profile is required. 
It remains to be seen whether large scale numerical simulations of zenith angle distributions of neutrinos show such crossings.

In this paper, we explore how these fast flavor conversions get affected if neutrinos have interactions beyond the Standard Model (SM), for example, with a new neutral gauge boson. Non-standard interactions of neutrinos with charged fermions are well constrained \cite{Guzzo:2004ue,Friedland:2004pp,Gonzalez-Garcia:2015qrr}, yet can give rise to interesting results inside a SN \cite{EstebanPretel:2007yu}. However, non-standard self-interactions (NSSI) of neutrinos are loosely constrained since they have not been directly observed yet \cite{Bardin:1970wq,Kolb:1987qy,Bilenky:1994ma,Masso:1994ww,Bilenky:1999dn,Belotsky:2001fb,Bakhti:2017jhm}. Such NSSI can give rise to an effective four-neutrino operator of the form $G_F \left(G^{\alpha\beta}\, \bar{\nu}_{{\rm L}\alpha}\gamma^\mu\nu_{{\rm L}\beta}\right)\,\left(G^{\zeta\eta}\,\bar{\nu}_{{\rm L}\zeta}\gamma_\mu\nu_{{\rm L}\eta}\right)$, where $\alpha=\beta$ indicates flavor-preserving NSSI (FP-NSSI), whereas $\alpha\neq\beta$ indicates flavor-violating NSSI (FV-NSSI). Such NSSI of neutrinos can affect collective oscillations.

The framework for studying the effects of FV-NSSI on self-induced oscillations was developed in \cite{Blennow:2008er}. While ordinary collective oscillations with SM interactions is known to be equivalent to an inverted pendulum in flavor space,  the addition of FV-NSSI acts like an external force on the pendulum. A subsequent detailed analysis, which included FV-NSSI and FP-NSSI \cite{Das:2017iuj}, demonstrated using box spectra and single-angle analysis that FP-NSSI leads to the pinching of spectral swaps, hence the suppression of collective oscillations, whereas FV-NSSI may lead to the development of swaps away from (or even in the absence of) a spectral crossing. In particular, FV-NSSI can give rise to collective oscillations during the neutronization burst epoch for both hierarchies, leading to distinct features in the neutronization spectra. 

Note that the above NSSI analyses were applicable in the absence of fast oscillations. This assumption was valid in the neutronization burst epoch, when only electron neutrinos are emitted and hence there cannot be collective effects or fast oscillations in SM. In this paper we carry out a detailed study of the effects of NSSI on fast as well as slow collective flavor conversions. 

The complete analytic solution to neutrino flavor conversions in a realistic situation involving a SN still remains intractable, and we have to resort to approximations to get insights into the complicated dynamics. In this paper, we work with the two-box spectrum (for the bipolar analysis), and the four-beam model (for the fast oscillation analysis). In the literature on SN neutrinos, these two simplified scenarios are considered to be ``the standard scenarios'' (see \cite{Hannestad:2006nj,Chakraborty:2016yeg,Raffelt:2007xt,Raffelt:2007cb,Banerjee:2011fj,Raffelt:2013rqa,Sawyer:2005jk,Sawyer:2008zs,Sawyer:2015dsa,Chakraborty:2016lct,Dasgupta:2016dbv,Sen:2017ogt,Izaguirre:2016gsx,Capozzi:2017gqd}), and provide ``proof of principle'' arguments for possible new effects.
 
We start by performing a linear stability analysis in the two-neutrino flavor space to analytically understand the effects of NSSI on the onset of collective oscillation. Such an analysis typically leads to an eigenvalue equation \cite{Banerjee:2011fj}, whose exponentially growing eigenvalues correspond to an instability, and indicate the onset. We find that with both FP-NSSI and FV-NSSI present, one also gets linearly increasing solutions in addition to the exponentially increasing ones. These linear solutions may lead to an earlier onset, and obviate the need for a seed to start collective oscillations. We demonstrate this using a two-box spectrum and a single emission angle, at distances far away from the neutrinosphere.

In order to analyze fast oscillations, we use the intersecting four-beam model of neutrinos and antineutrinos. We demonstrate that fast oscillations are suppressed by FP-NSSI even for rapidly growing temporal solutions, which would have been impossible otherwise \cite{Dasgupta:2015iia,Dasgupta:2016dbv}. On the other hand, FV-NSSI enhance fast oscillations and also cause them to start earlier. They also allow fast oscillations to take place even when the angle between the neutrino and the antineutrino is obtuse, which is not allowed in SM.

An important consequence of the last result above is that FV-NSSI can induce fast oscillations in two back-to-back beams of neutrino-antineutrino even when no spatial inhomogeneities are present. This would have been impossible in SM, where spatial inhomogeneities are necessary \cite{Chakraborty:2016lct}.

Finally, we study the effects of NSSI on the long-time flavor evolution of the four-beam model by solving the fully non-linear equations of motion numerically. We demonstrate that the fast oscillations are modulated by the slow oscillations. The frequency and amplitude of the modulations are influenced by the values of the NSSI parameters. 

We discuss these ideas in the following sections. In Section \ref{sec:FlavEvol}, we remind ourselves of the formalism and conventions for dealing with NSSI. In Section \ref{sec:slow}, we perform the linear stability analysis at distances much larger than the neutrinosphere, in the presence of NSSI. We further study the consequences for a simple box spectrum using a single-angle analysis. In Section \ref{sec:FastFlavConv}, we consider an intersecting four-beam model and explore the effects of FP-NSSI and FV-NSSI separately on fast oscillations. We also study the interplay between fast and slow oscillations. Finally, in Section \ref{sec:Conclusions}, we discuss our results and future prospects.

\section{The formalism}
\label{sec:FlavEvol}
We work in terms of the $2\times2$ flavor density matrices $\varrho_\mathbf{p}$, which are implicit functions of position ${\bf r}$ and time $t$. The diagonal entry of $\varrho_\mathbf{p}$ gives the probability for the particular flavor whereas the off-diagonal entries encode the phase information. 
The EoM for each momentum mode ${\bf p}$, in the absence of collisions, is given by \cite{Sigl:1992fn,Strack:2005ux}
\begin{equation}\label{eq:eom}
\partial_t \varrho_{{\bf p}} + {\bf v}_{\bf p} \cdot \nabla_{\bf x} \varrho_{{\bf p}}=- i [\mathcal{H}_{{\bf p}}, \varrho_{{\bf p}}]
\,\,,
\end{equation}
where ${\bf v}_{\bf p}$ is the velocity of the neutrino with momentum ${\bf p}$. The Hamiltonian matrix $\mathcal{H}_{{\bf p}}$ consists of the vacuum, matter, and self-interaction terms, and is given by 
\begin{equation}
\mathcal{H}_{{\bf p}}= \mathcal{H}^{{\rm vac}}_{\bf p} + \mathcal{H}^{\rm MSW} + \mathcal{H}^{\nu\nu}_{\bf p}\,\ .
\label{eq:ham}
\end{equation}
Our convention is such that it allows us to consider neutrinos and antineutrinos on the same footing with the vacuum term getting a negative sign for antineutrinos. In other words, the antineutrino spectrum may be written as a continuation of the neutrino spectrum to negative energies.

In the flavor basis, the vacuum term is given by
\begin{equation}
\mathcal{H}^{{\rm vac}}_{\bf p}= \mathcal{U}~\textrm{diag}(-\omega/2, +\omega/2)~\mathcal{U}^\dagger \, ,
\end{equation}
 where $\omega\equiv(m_2^2-m_1^2)/(2E)$ and $E=|{\bf p}|$ for ultra-relativistic neutrinos. Here $m_i$ is the mass of the neutrino mass eigenstate $\nu_i$. The unitary $2\times2$ matrix $\mathcal{U}$ is given by
 \begin{equation}
  \mathcal{U}=\begin{pmatrix}
               \cos\vartheta_0 & -\sin\vartheta_0\\
               \sin\vartheta_0 & \cos\vartheta_0 
              \end{pmatrix}\,,\nonumber
 \end{equation}
where $\vartheta_0$ is the mixing angle.
The MSW potential term, due to charged current interactions with the background 
electron density $n_e$, is represented by
\begin{equation}
\mathcal{H}^{\rm MSW}=  \sqrt{2} G_F n_e\,\ \textrm{diag} (1,0) \,. 
\end{equation}
Finally, the most general effective Hamiltonian due to neutrino self-interactions is given by \cite{Sigl:1992fn}
\begin{eqnarray}\label{self}
\mathcal{H}^{\nu\nu}_{\bf p} &=&\sqrt{2}G_F  \int \frac{d^3 {\bf q}}{(2 \pi)^3}  (1 -{\bf v}_{\bf p}\cdot {\bf v}_{\bf q})\times\nonumber\\
                        & &\bigl\{G({\varrho_{\bf q}} - {\bar\varrho_{\bf q}})G 
                        +G~{\rm Tr}\left[({\varrho_{\bf q}} - {\bar\varrho_{\bf q}}) G\right] \bigr\} \,\ ,
\end{eqnarray}
where the term $(1 -{\bf v}_{\bf p}\cdot {\bf v}_{\bf q})$ gives rise to multi-angle effects due to neutrinos travelling on different trajectories.
Here $G$ is a dimensionless coupling matrix, which is equal to the identity matrix within the SM.

A comment about some of the notations in this article is desirable.
A careful analysis without deviating much from the notations in the earlier literature needs the introduction of many similar-looking symbols. 
For clarity, Appendix \ref{glossary} gives a table with the list of such symbols and their meanings.
 \bigskip

\subsection{Introducing NSSI parameters}
After including NSSI, the coupling matrix $G$ in Eq.\,(\ref{self}) becomes \cite{Das:2017iuj}
\begin{equation}\label{coupling}
 G=\begin{bmatrix}
      1+\gamma_{ee} & \gamma_{ex} \\
      \gamma_{ex}^* & 1+\gamma_{xx}
      \end{bmatrix}.
\end{equation}
Processes involving neutrino self-interactions are very rare and hence difficult to constrain directly. However, direct constraints are available from flavor physics \cite{Bardin:1970wq,Bakhti:2017jhm}, supernova cooling bounds \cite{Kolb:1987qy} and invisible $Z$-width data from LEP \cite{Bilenky:1994ma,Bilenky:1999dn}. Taking these into account, the constraints on NSSI mediated by a gauge boson translate to $ |\gamma_{ee}|,\, |\gamma_{xx}|\,{\rm and}\,|\gamma_{ex}|\sim\, \mathcal{O}(1)$ \cite{Das:2017iuj}. Indirect constraints arising from the bounds on neutrino interactions with charged fermions \cite{Guzzo:2004ue,Friedland:2004pp,Gonzalez-Garcia:2015qrr} are somewhat stronger. Note that stringent bounds can be imposed on neutrino NSSI from SU(2)$_L$ gauge invariance \cite{Gavela:2008ra}. However, these can be evaded in certain models, where active neutrinos mix with new Dirac fermions charged under a U$(1)'$ gauge group \cite{Farzan:2016wym}. In this paper, the couplings are restricted to $\mathcal{O}(0.01-0.1)$.

Following \cite{Blennow:2008er}, one can also write the coupling matrix in the Pauli basis as 
\begin{equation}\label{GG}
G=\frac{1}{2}\left(g_0{\mathbb I}+{\boldsymbol g}\cdot {\bf\sigma}\right) \,,
\end{equation}
such that $g = \{g_0, {\boldsymbol g}\}$ represents the net neutrino-neutrino coupling.
From the above equations, one arrives at $g_0=2+\gamma_{ee}+\gamma_{xx},\,g_1=2\,{\rm Re}(\gamma_{ex}),\,g_2=2\,{\rm Im}(\gamma_{ex}^*)$ and $g_3=\gamma_{ee}-\gamma_{xx}$. Hence, $g_0$ and $g_3$ are FP-NSSI couplings whereas $g_1$ and $g_2$ represent FV-NSSI couplings.

As shown in \cite{Das:2017iuj}, the parameter $g_0$ can be scaled away using the redefinitions
\begin{equation}\label{rescale}
  {\boldsymbol g} \to {\boldsymbol g}/(g_0/2)\,, \qquad \mu_{\text{\tiny R}} \to \mu_{\text{\tiny R}} (g_0/2)^2\,,
\end{equation}
where $\mu_{\text{\tiny R}}$ will be defined presently.
Further simplification can be achieved by redefining the phase of $\nu_x$ such that $g_2=0$.
This allows us to write the redefined coupling matrix as 
\begin{equation}\label{redefcoupling}
 G=\begin{bmatrix}
      1+g_3 & g_1 \\
      g_1 & 1-g_3
      \end{bmatrix}\,.
\end{equation}
Henceforth, we will work with this coupling matrix.

\subsection{Setting up the problem}
We confine ourselves to a spherically symmetric setup, where neutrinos are emitted from a fiducial neutrinosphere of radius $R$. Following \cite{Banerjee:2011fj},
we label them by the variable $u=\sin^2\vartheta_R$, where $\vartheta_R$ is the emission angle of the neutrinos. For simplicity, we assume that the solution is stationary and has an axial symmetry. The radial velocity for a mode $u$ at the radius $r$ is $v_{r,u}=\sqrt{1-u R^2/r^2}$ .

In terms of the flux matrices $F$ \cite{Banerjee:2011fj,Raffelt:2013rqa}
\begin{equation}
F_{\w,u}d\w\,du=2\pi r^2\,v_{r,u}\varrho_{{\bf p}} \frac{d^3{\bf p}}{(2\pi)^3}\,,
\end{equation}
the EoMs become 
\begin{equation}
i \partial_r F_{\w,u} =[H_{\w,u},F_{\w,u}]\,,
\end{equation}
where 
\begin{eqnarray}\label{newHam}
H_{\w,u} &=&(\w+\lam_r)v_{r,u}^{-1} +\mu_{\text{\tiny R}} \frac{R^2}{r^2} \int\,d\Gamma'\frac{1-v_{r,u}v'_{r,u'}}{v_{r,u}v'_{r,u'}}\times\nonumber\\
        & &\left\{G F_{\w,u}G + G~{\rm Tr}\left[F_{\w,u}G\right] \right\} \,.
\end{eqnarray}
Here $\int\,d\Gamma'=\int_{-\infty}^{\infty}\,d\w'\int_{0}^{1}\,du'$, and negative values of $\w$ represent antineutrinos. The quantities $\lam_r$ (matter potential at a radius $r$) and $\mu_{\text{\tiny R}}$ (neutrino-neutrino potential at the neutrinosphere) are defined as 
\begin{eqnarray}
 \lam_r&=& \sqrt{2}G_F\,n_e(r)\,\,,\nonumber\\
 \mu_{\text{\tiny R}} &=& \frac{\sqrt{2}G_F\,\left[F_{\w,u}^{\bar{e}}(R)-F_{\w,u}^{\bar{x}}(R)\right]}{4\pi R^2}\,,
\end{eqnarray}
where $F_{\w,u}^{\bar{e}}(r)$ represents the $\bar{\nu}_e\bar{\nu}_e$ flavor-diagonal element of the $2\times2$ matrix $F_{\w,u}(r)$, at a radius $r$, for $\w<0$, i.e., for antineutrinos.
In our analysis, $\mu_{\text{\tiny R}}$ is further rescaled as given in Eq.\,(\ref{rescale}).

The flux matrices $F_{\w,u}$ in Eq.\,(\ref{newHam}) have been rescaled such that at $t=0$,
\begin{equation}
 \int\,d\Gamma\,\left[F_{\w,u}^{\bar{e}}(R)-F_{\w,u}^{\bar{x}}(R)\right]=1\,.\nonumber
\end{equation}
These $F_{\w,u}$ may now be written in the form
\begin{equation}\label{Fdef}
F_{\w,u}=\frac{{\rm Tr\,}(\,F_{\w,u})}{2} +\frac{g_{\w,u}}{2}\begin{pmatrix}
                                                \,s_{\w,u} && S_{\w,u}\\
                                                S_{\w,u}^* && -s_{\w,u}
                                               \end{pmatrix}\,,
\end{equation}
where 
\begin{equation}\label{Diffspectrum}
g_{\w,u}= \left\{\begin{array}{lr}
    F^e_{\w,u}-F^x_{\w,u} & {\rm for~ }\w>0\,~~,\\
    F_{\w,u}^{\bar{x}}-F_{\w,u}^{\bar{e}}& {\rm for~}\w<0\,~~.
   \end{array}\right.
\end{equation}
is the difference in spectra of the two flavors and $S_{\w,u}$ is the off-diagonal parameter that we will use to characterize flavor conversions. In this entire analysis, we neglect all collisional processes which change the total number of neutrinos. Hence ${\rm Tr}\,(F_{\w,u})$ is conserved and can be dropped from the EoMs.

\section{Stability analysis with NSSI}
\label{sec:slow}

At $t=0$, we have $s_{\w,u}=1$ and $S_{\w,u}=0$ in Eq.\,(\ref{Fdef}).
As flavor evolution begins, $S_{\w,u}$ starts developing a non-zero value. Since $s_{\w,u}^2+S_{\w,u}^2=1$, a small amplitude expansion may be performed with the approximation $s_{\w,u}\approx 1$, $S_{\w,u}\ll 1$, and where terms of $\mathcal{O}(S^2)$ are dropped. This is equivalent to linearizing the equations in $S_{\w,u}$.

The linearized EoMs are
\begin{eqnarray}\label{stability-eom}
i\partial_r\,S_{\w,u}&=&\biggl[\left(\w+\lam_r\right)v_{u,r}^{-1} \nonumber\\ & &\,+\mu_{\text{\tiny R}}\,\frac{R^2}{r^2}\,(1-g_1^2+3g_3^2+4g_3)\times\nonumber\\
                                    & &\phantom{++}\int\,d\Gamma'\,\frac{1-v_{u,r}v_{u',r'}}{v_{u,r}v_{u',r'}}g_{\omega',u'}\biggr]S_{\w,u}\nonumber\\
                                    &-&\mu_{\text{\tiny R}}\,\frac{R^2}{r^2}\,\int\,d\Gamma'\,\frac{(1-v_{u,r}v_{u',r'})}{v_{u,r}v_{u',r'}}g_{\omega',u'}\times\nonumber\\
                                    & &\biggl[(1+g_1^2-g_3^2)S_{\w',u'}+2g_1^2S_{\w',u'}^*+4\,g_1g_3\biggr]\,.\nonumber\\
\end{eqnarray}

Eq.\,(\ref{stability-eom}) clearly is not an eigenvalue equation, as it would have been in the SM limit \cite{Banerjee:2011fj,Raffelt:2013rqa}. This would lead to the following interesting consequences. (While describing these observations, we will drop the subscripts for simplicity of notation.)

(i) Only FP-NSSI: In this limit, Eq.\,(\ref{stability-eom}) is an eigenvalue equation and the standard analysis of \cite{Banerjee:2011fj} holds. One can look for exponentially growing solutions of the form $S=Q e^{-i \Omega t}$, where $\Omega=\gamma+i\kappa$ is complex. A positive non-zero value of $\kappa$ indicates an instability growing with a rate $e^{\kappa\,t}$.

(ii) Only FV-NSSI: The EoMs governing $S$ are not simple
eigenvalue equations in $S$ anymore. However one may combine the pair of coupled differential equations for $S$ and $S^*$ to get an eigenvalue equation.
This may be done by looking for solutions of the form $S=A e^{ \Gamma t}$, where $A$ can be complex and $\Gamma$ is real. Positive solutions of $\Gamma$ indicate a runaway solution and hence signal an instability.

(iii) Both FP-NSSI and FV-NSSI: Eq.\,(\ref{stability-eom}) cannot be converted to a simple eigenvalue equation. The term proportional to $g_1g_3$ would generate $S$ even if it were vanishing at $t=0$. As long as  $S$ is sufficiently small, the growth rate will be dominated by a linear rise owing to this term. However, as $S$ grows, the exponential growth may take over. Hence one expects to find a linear rise, followed by an exponential one in the instability growth rates.

In the next section, we shall demonstrate these observations explicitly using a simple two-box spectrum in the single-angle approximation and far away from the neutrinosphere. This will also provide an analytical understanding of the numerical results presented in \cite{Das:2017iuj}.

\subsection{Analytical understanding of the evolution of a two-box spectrum}
In \cite{Das:2017iuj}, it was shown that in the single-angle approximation and far away from the neutrinosphere, FP-NSSI can cause suppression of collective oscillations, leading to pinching of spectral swaps in a two-box spectrum. Conversely, presence of FV-NSSI leads to a gradual widening of spectral swaps. We will now try to explain these observations analytically using the formalism developed in the earlier section.

Far away from the neutrinosphere $(r\gg R)$, we can drop terms of $\mathcal{O}(R^2/r^2)$ in Eq.\,(\ref{stability-eom}). In this limit, the EoMs are given by
\begin{eqnarray}\label{stability-eomRr}
 i\partial_r S_{\w,u}&=&\left[\omega+\lam_r+u\widetilde{\lam}_r+u\widetilde{\mu}_r\epsilon\,(1-g_1^2+3g_3^2+4g_3)\right]S_{\w, u}\nonumber\\
&-&\widetilde{\mu}_r\,(1-g_3^2+g_1^2) \int du'\,d\omega'\,(u+u')\,g_{\omega',u'}\,S_{\w',u'}\,\nonumber\\
&-&2\widetilde{\mu}_r\, g_1^2 \int du'\,d\omega'\,(u+u')\,g_{\omega',u'}\,S_{\w',u'}^{*}\,\nonumber\\
&-&4\widetilde{\mu}_r\, g_1 g_3 \int du'\,d\omega'\,(u+u')\,g_{\omega',u'}\,,
\end{eqnarray}
where
\begin{eqnarray}
\widetilde{\lam}_r&=& \sqrt{2}G_F\,n_e(r)\,\frac{R^2}{2\,r^2}\,,\nonumber\\
\widetilde{\mu}_r &=& \frac{\sqrt{2}G_F\,}{4\pi R^2}\,\frac{R^4}{2\,r^4}\,\left[F_{\w,u}^{\bar{e}}(R)-F_{\w,u}^{\bar{x}}(R)\right]\,.
\end{eqnarray}
Here $\epsilon=\int\,d\w'du'g_{\w',u'}$ encodes the net neutrino-antineutrino asymmetry.
This is the generalization of the multi-angle evolution equation derived for the SM, and reduces to it in the limit $g_1,g_3\rightarrow0$ [Eq.\,(31) in \cite{Banerjee:2011fj}]. 

\begin{figure}[!t]
\includegraphics[scale=0.4]{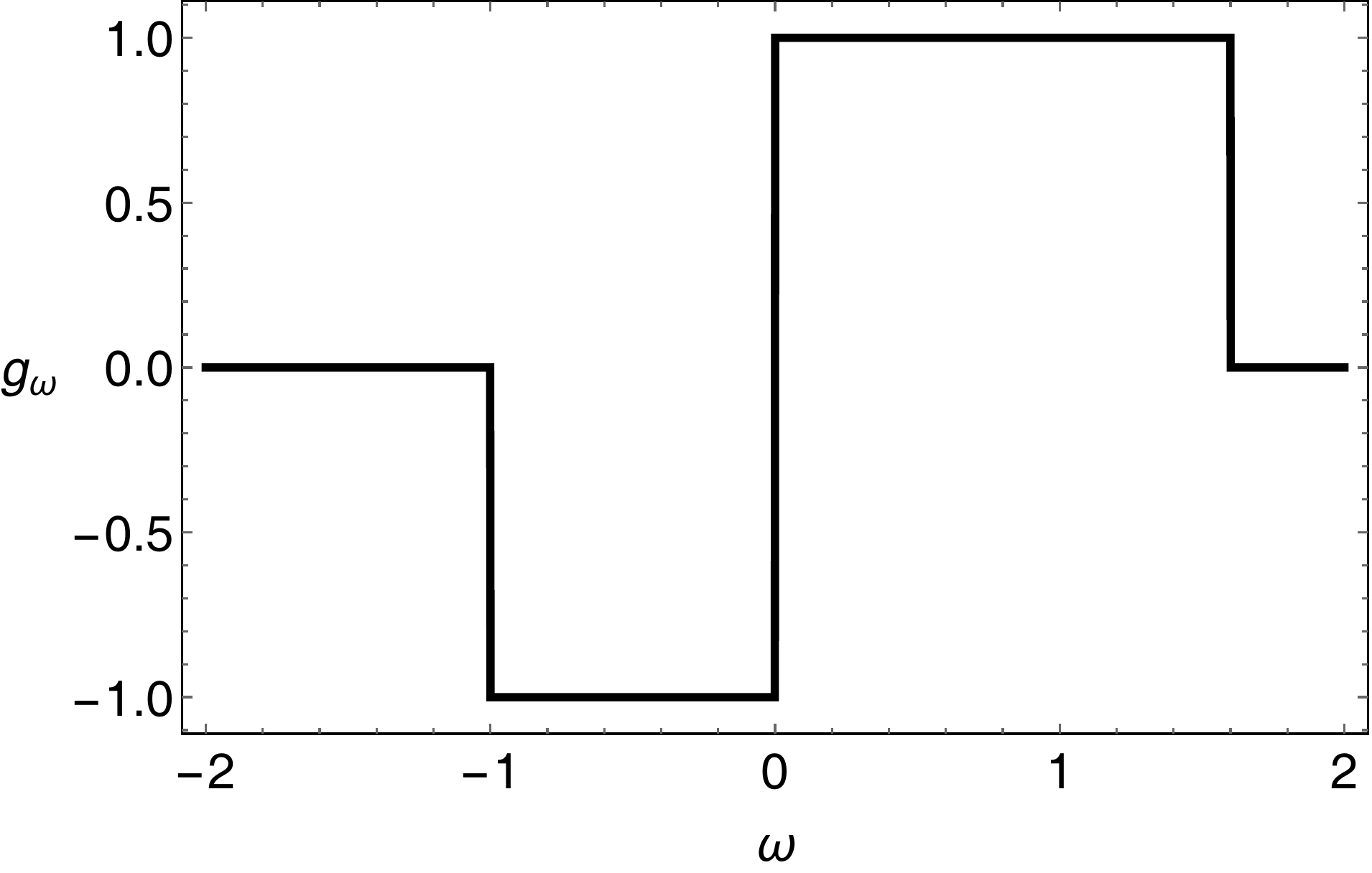}\\ \includegraphics[width=8cm,height=5cm]{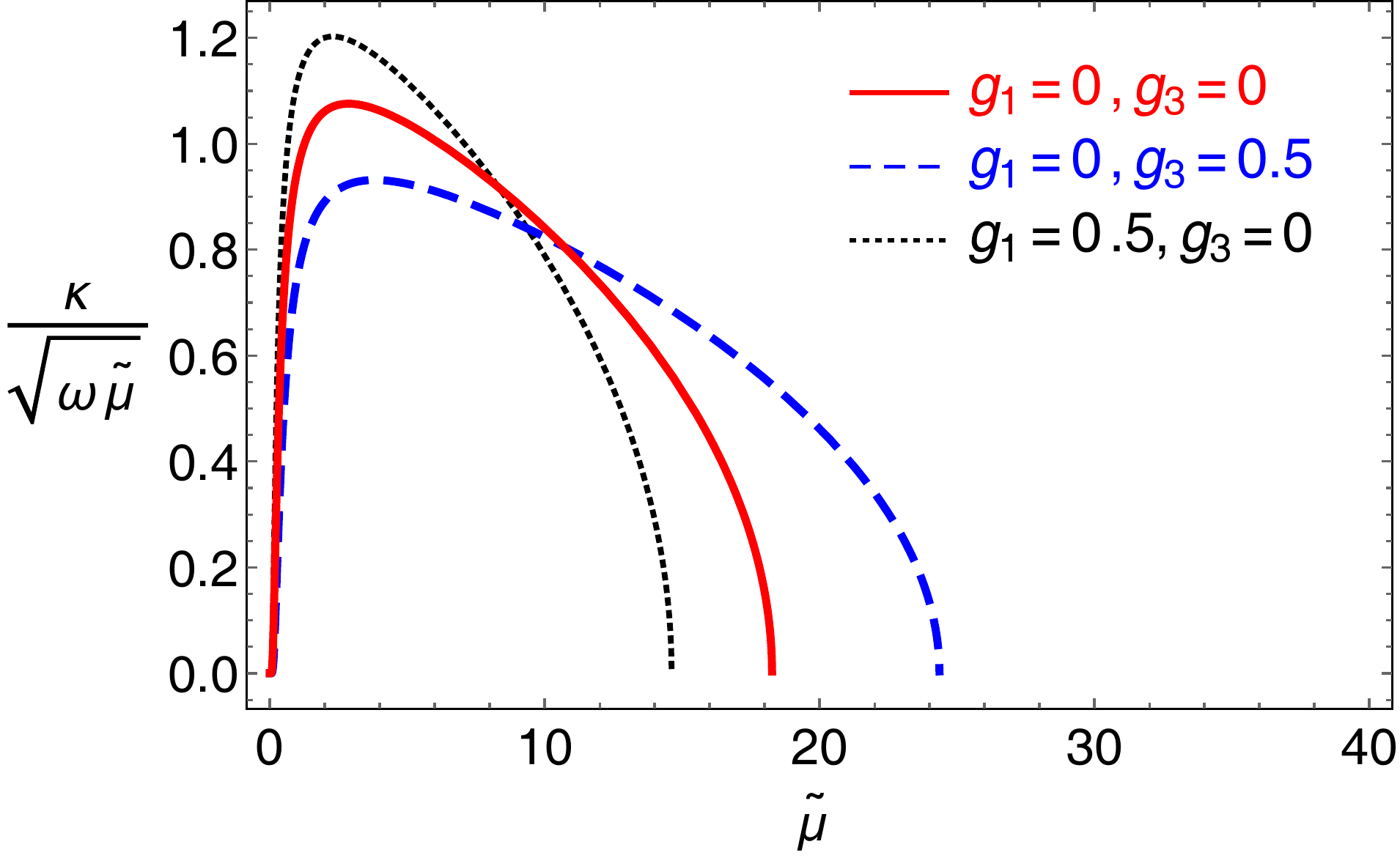}
\caption{Top panel: a two-box spectrum [Eq.\,(\ref{twobox})] with $A=1$ and $B=1.6$. Bottom panel: Plot showing growth rates $\kappa$ in units of $\sqrt{\w\widetilde{\mu}}$ for $u=1/2$. Red (solid) indicates the SM, whereas blue (dashed) represents $g_1=0\,,g_3=0.5$ and black (dotted) shows $g_1=0.5\,,g_3=0$.}
\label{fig0}
\end{figure}

For demonstrating the effects of NSSI, we consider the scenario where all the neutrinos have the same emission angle and hence may be labelled by a single angular mode $u_0$. We take the initial spectrum to be the two-box spectrum, as shown in the top panel of Fig.\,\ref{fig0}:
\begin{equation}\label{twobox}
 g_\w\equiv g_{\w,u_0}=\begin{cases}
   -1 & -A<\w<0\,\,,\\
   +1 & ~~0<\w<B\,\,.
  \end{cases}
\end{equation}
Such a simple box spectrum has the advantage of making the eigenvalues analytically tractable and hence the effects of NSSI become clear.
Following \cite{Banerjee:2011fj}, one can define
\begin{equation}
 \overline{\lam}_r\equiv\lam_r+u\bigl[\widetilde{\lam}_r+\widetilde{\mu}_r\,\epsilon\,(1-g_1^2+3g_3^2+4g_3)\bigr]\,,
\end{equation}
 which acts as the effective matter term in the equations. In the following discussions in this section, we drop the subscript $r$ of $\widetilde{\mu}$ for simplicity of notation.

The suppression and the enhancement of collective oscillations can be characterized in terms of the change of the growth rate $\kappa$.
The growth rates in this case are proportional to $\sqrt{\w\widetilde{\mu}}$, and hence come under slow collective oscillations. The bottom panel of Fig.\,\ref{fig0} shows the growth rates in units of $\sqrt{\w\widetilde{\mu}}$ for $u=u_0=1/2$. For any other value of $u_0$, the results will be identical with $\widetilde{\mu}$ replaced by $2\,u_0\,\widetilde{\mu}$.
The effects of NSSI may be observed and interpreted as follows:

 (i) Only FP-NSSI: As noted before, in this case Eq.\,(\ref{stability-eomRr}) is a simple eigenvalue equation. The effective matter term can be rotated away by going to the appropriate co-rotating frame. The analytical results for SM [see Eq.\,(47) of \cite{Banerjee:2011fj}] then carry through with $\widetilde{\mu}$ replaced by $\widetilde{\mu}(1-g_3^2)$, and hence the collective oscillations are suppressed due to $g_3$. 

 (ii) Only FV-NSSI: In this scenario, Eq.\,(\ref{stability-eomRr}) is not a simple eigenvalue equation. However, if the term $g_1^2S_{\w,u}^*$ can be neglected when compared to $(1+g_1^2)S_{\w,u}$, then Eq.\,(\ref{stability-eomRr}) may be approximated by an eigenvalue equation, and the effective matter term can be co-rotated away. The numerical observation of enhancement of collective oscillations due to $g_1$ may be then qualitatively interpreted as a result of the $\widetilde{\mu}(1+g_1^2)$ factor. The effect of the neglected $g_1^2S_{\w,u}^*$ term is difficult to determine analytically, however.
 
 (iii) When both FP-NSSI and FV-NSSI are present, Eq.\,(\ref{stability-eomRr}) cannot be converted to an eigenvalue equation and hence an analytical understanding in terms of linear stability analysis seems elusive.

\section{Fast flavor oscillations: the four-beam model}
\label{sec:FastFlavConv}

\begin{figure}[!t]
\includegraphics[scale=0.3]{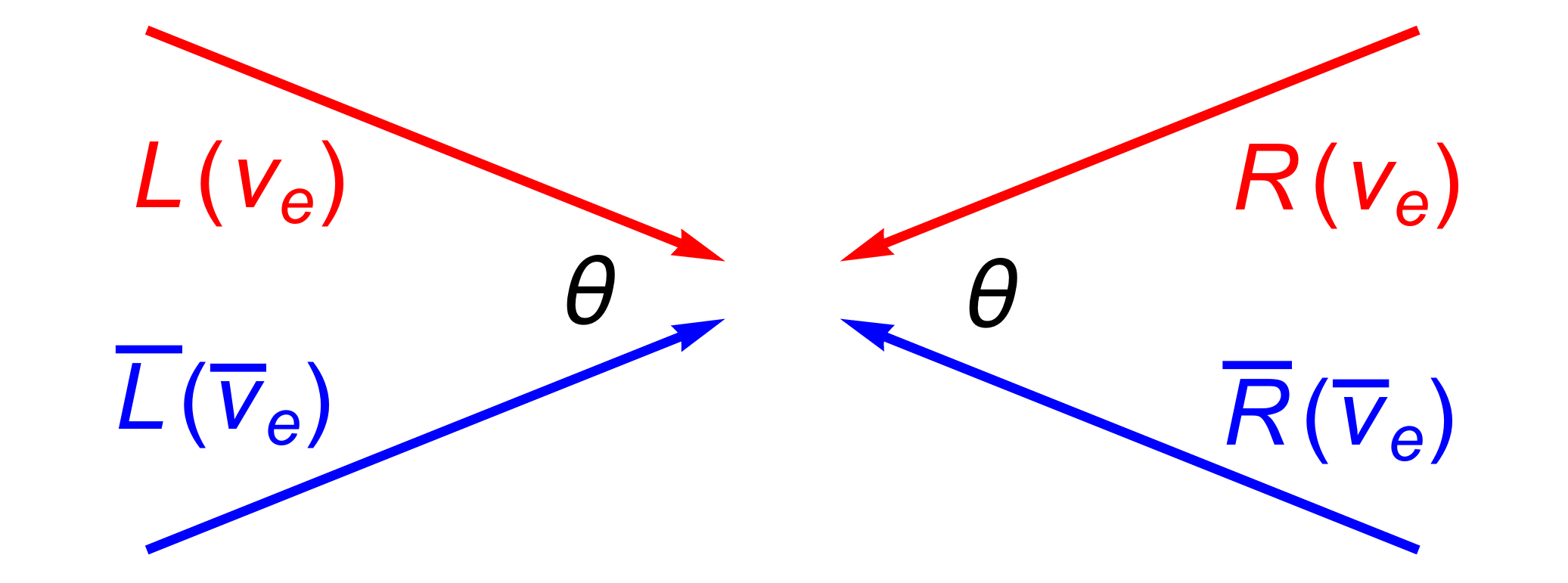}
\caption{Intersecting Four Beam Model.}
\label{fig1}
\end{figure}

In this section, we demonstrate the effects of neutrino NSSI on a simple system that shows fast flavor oscillations, an intersecting four-beam model consisting of two right-going and left-going neutrinos and  antineutrinos each \cite{Chakraborty:2016lct}, as shown in Fig.\,\ref{fig1}. Since our understanding of the fast oscillations phenomenon is still in an exploratory phase, we have considered the simplest system showing such an effect \cite{Chakraborty:2016lct,Izaguirre:2016gsx,Capozzi:2017gqd}, to analytically understand the phenomenon without delving deeper into a more realistic spectra.

Following \cite{Chakraborty:2016lct}, the amplitudes for the corresponding modes are denoted by $Q_L$ $(Q_{\bar{L}})$ for neutrinos (antineutrinos) coming from left, and $Q_R$ $(Q_{\bar{R}})$ for neutrinos (antineutrinos) coming from right. Their corresponding spectra are $g_{\text{\tiny L}},g_{\bar{\text{\tiny L}}}$,
$g_{\text{\tiny R}}$ and $g_{\bar{\text{\tiny R}}}$. The spectra here are taken to be left-right symmetric, i.e,
\begin{eqnarray}
 g_{\text{\tiny R}}=g_{\text{\tiny L}}&=&\frac{1}{2}(1+a)\nonumber\,,\\
 g_{\bar{\text{\tiny R}}}=g_{\bar{\text{\tiny L}}}&=&-\frac{1}{2}(1-a)\,,
\end{eqnarray}
where $a$ gives the net neutrino-antineutrino asymmetry in the system. The range of $a$ is chosen to be $-1\leq a \leq 1$. The angle at the intersection of the two beams is denoted by $\theta$ as shown in the figure.

\begin{figure*}
\includegraphics[scale=0.4]{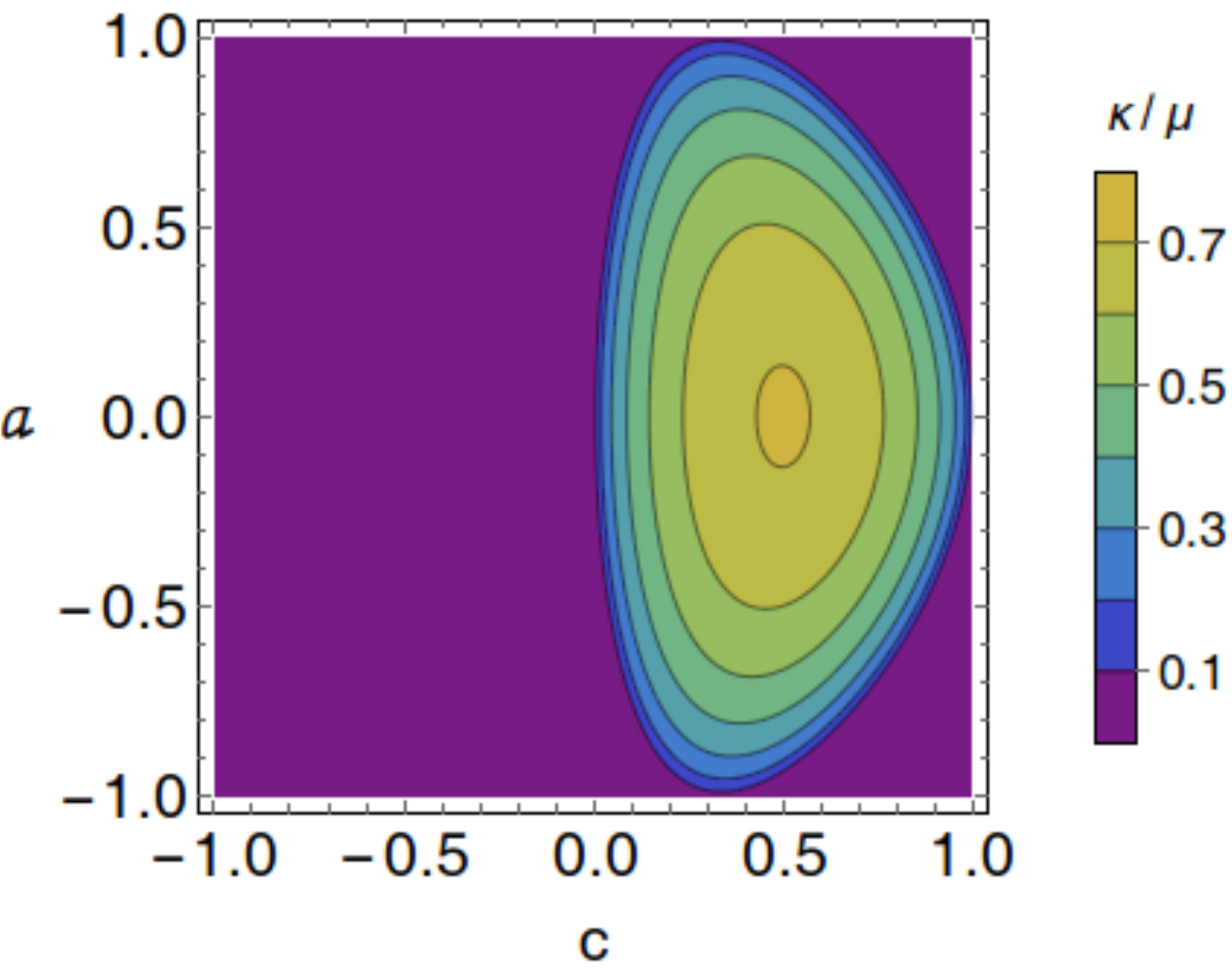}~~\includegraphics[scale=0.4]{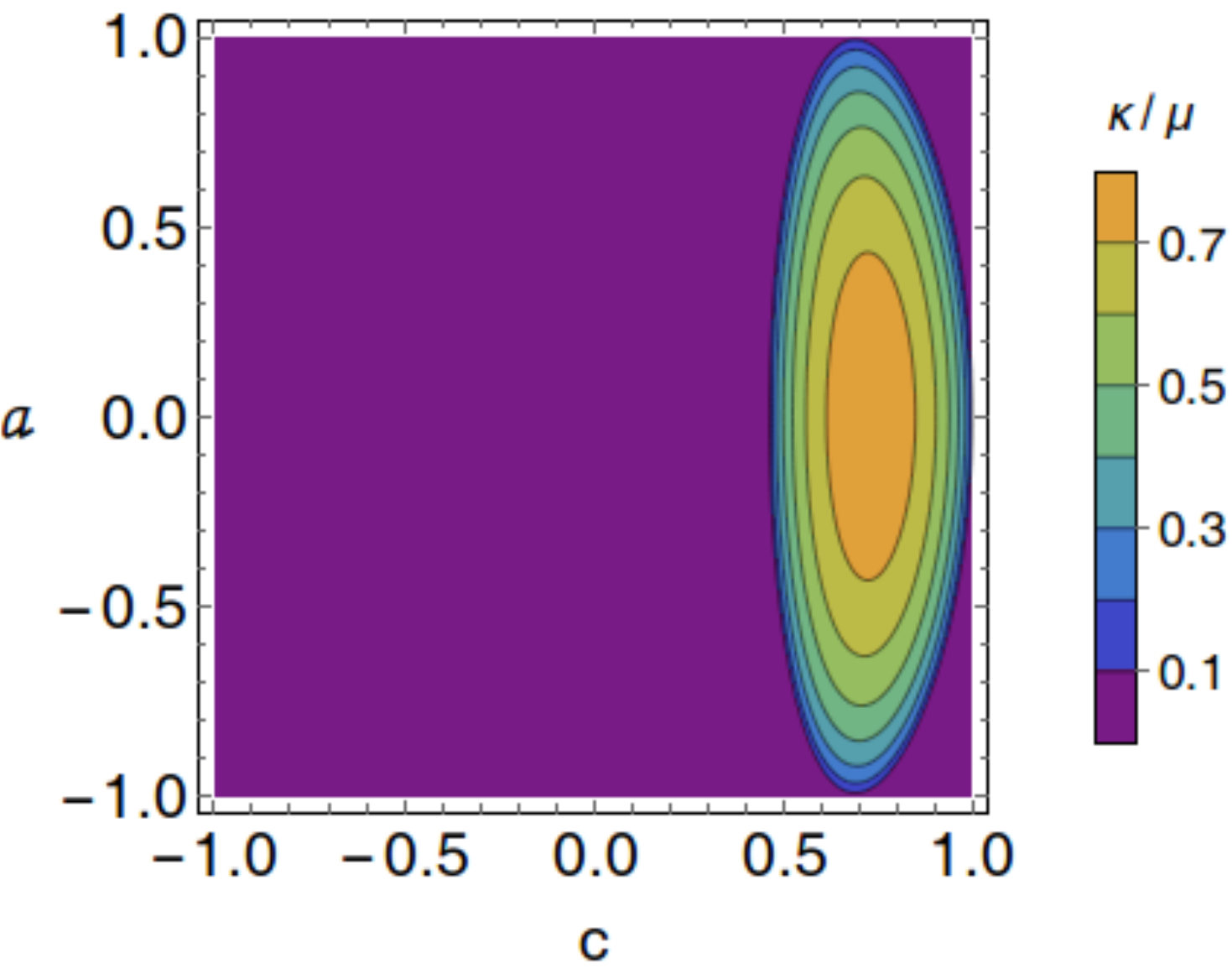}~~\includegraphics[scale=0.4]{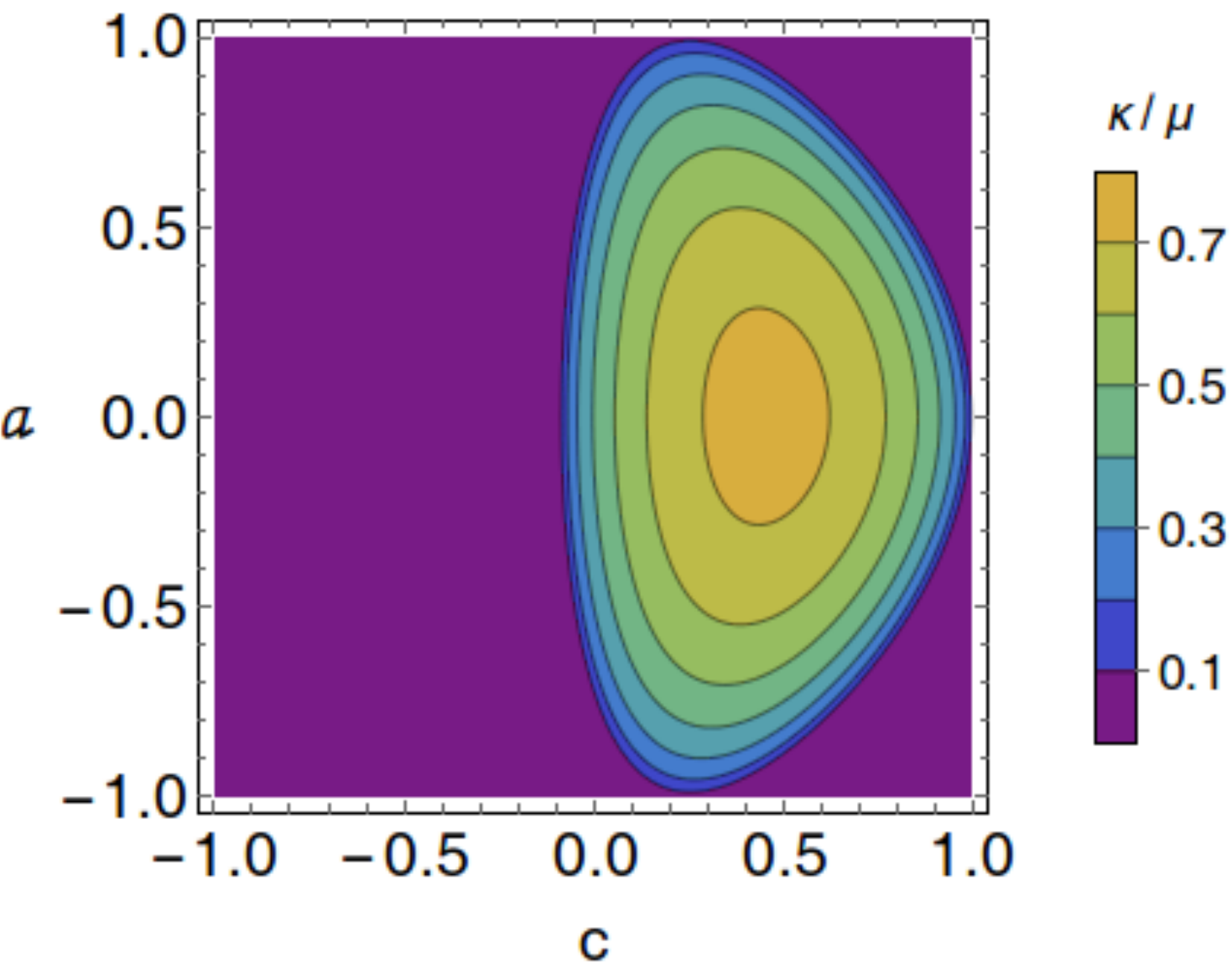}
\caption{Dependence of growth rates on $c\equiv\cos\theta$ and the $\nu-\bar{\nu}$ asymmetry $a$, for the left-right symmetry-breaking solution $(Q_-,\bar{Q}_-)$. Left: $g_1,\,g_3=0$, Centre: $g_1=0\,,g_3=0.3$, Right: $g_1=0.3\,,g_3=0$.}
\label{fig2}
\end{figure*} 

Since we are interested in conversions taking place very close to the neutrinosphere, the effective neutrino-neutrino potential $\mu\approx \mu_{\text{\tiny R}}$, and the effective matter potential $\lam\approx\lam_{\text{\tiny R}}=\sqrt{2} G_F n_e(R)$. Also, it is more useful to label the modes using the velocity vector ${\bf v}$ rather than the variable $u=\sin^2\vartheta_R$. 
The EoM for the off-diagonal parameter $S_{\bf p}$ for each mode ${\bf p}$ is given by
\begin{eqnarray}\label{fastEoM}
i(\partial_t&+&{\bf v_{\bf p}}\cdot\nabla)S_{\bf p}\nonumber\\& &
 =\biggl[\w+\lam+\mu\,(1-g_1^2+3g_3^2+4g_3)\times\nonumber\\& & \phantom{spa}
                                     \sum_{{\bf q}}(1-{\bf v}_{\bf p}\cdot{\bf v}_{\bf q})g_{{\bf q}}\biggr]S_{\bf p} \nonumber\\
                                    & &-\mu\,\sum_{{\bf q}}(1-{\bf v}_{\bf p}\cdot{\bf v}_{\bf q})g_{{\bf q}}\times\nonumber\\
                                    & &\phantom{spa}\biggl[S_{\bf q}+(g_1^2-g_3^2)S_{\bf q}+2g_1^2S_{\bf q}^*+4\,g_1g_3\biggr]\,,~~\,~~~
\end{eqnarray}
where ${\bf q}$ stands for the other three modes.

In the four-beam model, $\lambda=0$. For the linear stability analysis, we look for exponentially growing solutions of the form
$S_q=Q_q e^{-i \Omega t}$, where $\Omega=\gamma+i\,\kappa$ is complex. A positive non-zero $\kappa$ indicates an instability in the system and $\kappa\sim\mu$ indicates fast oscillations.

\subsection{Linearized analysis of the model}

The symmetry of the intersecting four-beam model can be used to combine the neutrino-antineutrino amplitudes into $Q_\pm\equiv(Q_L\pm Q_R)/2$ and $\bar{Q}_\pm\equiv(Q_{\bar{L}}\pm Q_{\bar{R}})/2$. This allows us to decouple
the equations for four modes into two sets of two. Within the SM, the first set consists of $(Q_+,\bar{Q}_+)$, which is the left-right symmetric solution that undergoes slow collective oscillations, and the second set consists of $(Q_-,\bar{Q}_-)$, which is the left-right symmetry-breaking solution that undergoes fast oscillations \cite{Chakraborty:2016lct}. 

We find that the narrative of fast oscillations changes significantly in the presence of NSSI. 

(i) For the symmetry breaking solution, FV-NSSI (FP-NSSI) increases (decreases) the available parameter space for fast oscillations. In fact, for FV-NSSI,
oscillations can happen even for $\cos\theta\,<0$ which was not possible in the SM. 

(ii) Within the SM, fast oscillations are possible for $(Q_+,\bar{Q}_+)$ only if spatial homogeneity of the beam is broken. However, the presence of FV-NSSI allows for fast oscillations even for homogeneous beams. 

We now demonstrate the above features in the context of linear stability analysis. We work in the approximation where only either $g_1$ or $g_3$ is non-zero, and the $g_1^2S_{\bf q}^*$ from Eq.\,(\ref{fastEoM}) can be neglected. We have numerically checked that the latter is a good approximation for $g_1\lesssim\mathcal{O}(0.5)$. It is now possible to understand the above features analytically, by writing down eigenvalue equations in the form
\begin{equation}\label{eigeqn}
 \Omega\begin{pmatrix}
        Q_\pm\\
        \bar{Q}_\pm
       \end{pmatrix}=\begin{pmatrix}
                      \mathcal{H}_{11} && \mathcal{H}_{12}\\
                      \mathcal{H}_{21} && \mathcal{H}_{22}\\
                      \end{pmatrix}\begin{pmatrix}
                                          Q_\pm\\
                                          \bar{Q}_\pm
                                          \end{pmatrix}\,,
\end{equation}
where $\mathcal{H}_{ij}$'s and the corresponding eigenvalues are given in Appendix~\ref{app}. Note that in order to isolate fast oscillations, the eigenvalues are calculated in the limit $\w/\mu\rightarrow0$. This also automatically makes the analysis energy independent. 

\begin{figure}[!h]
\includegraphics[scale=0.4]{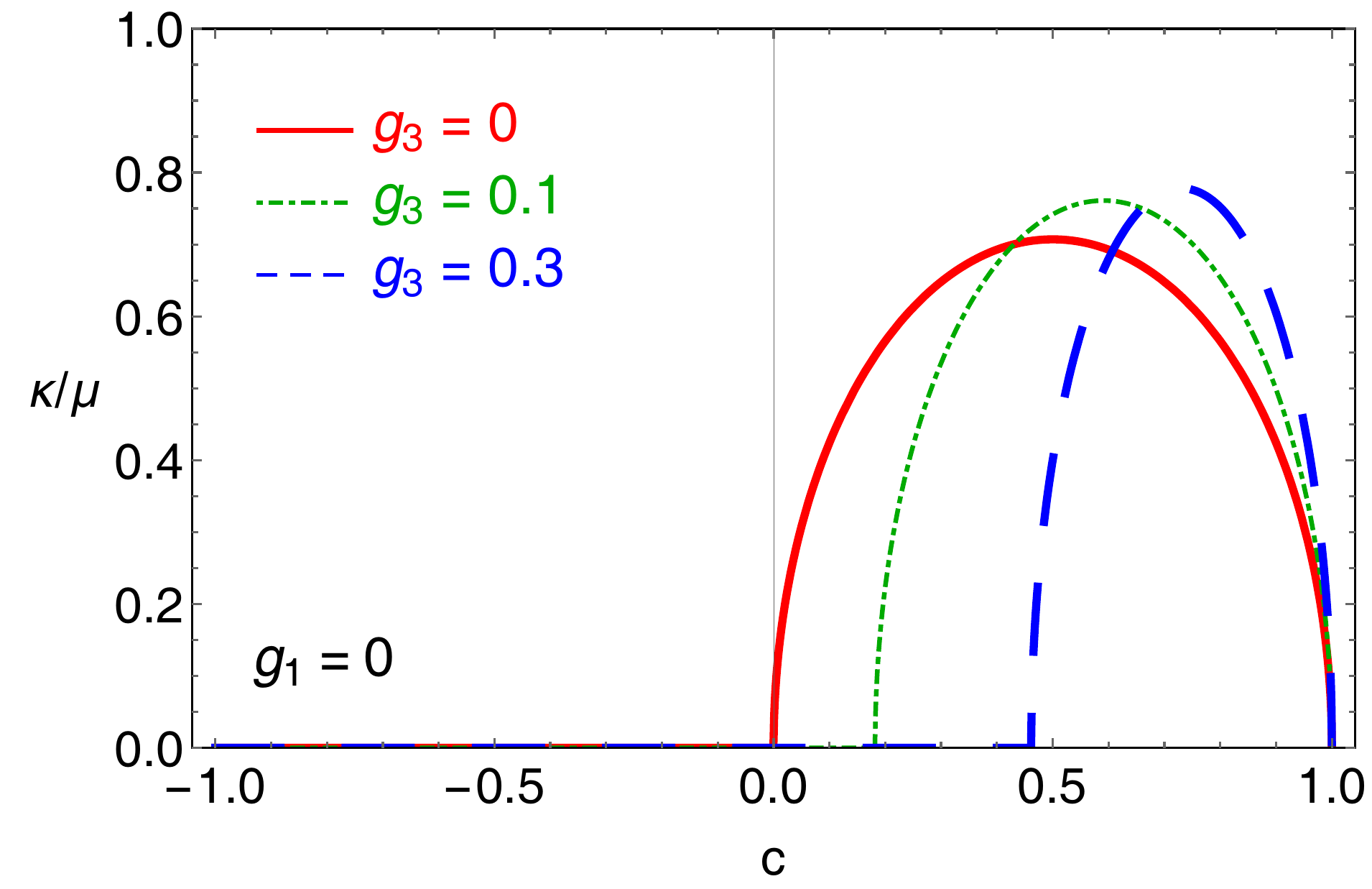}\\ \includegraphics[scale=0.4]{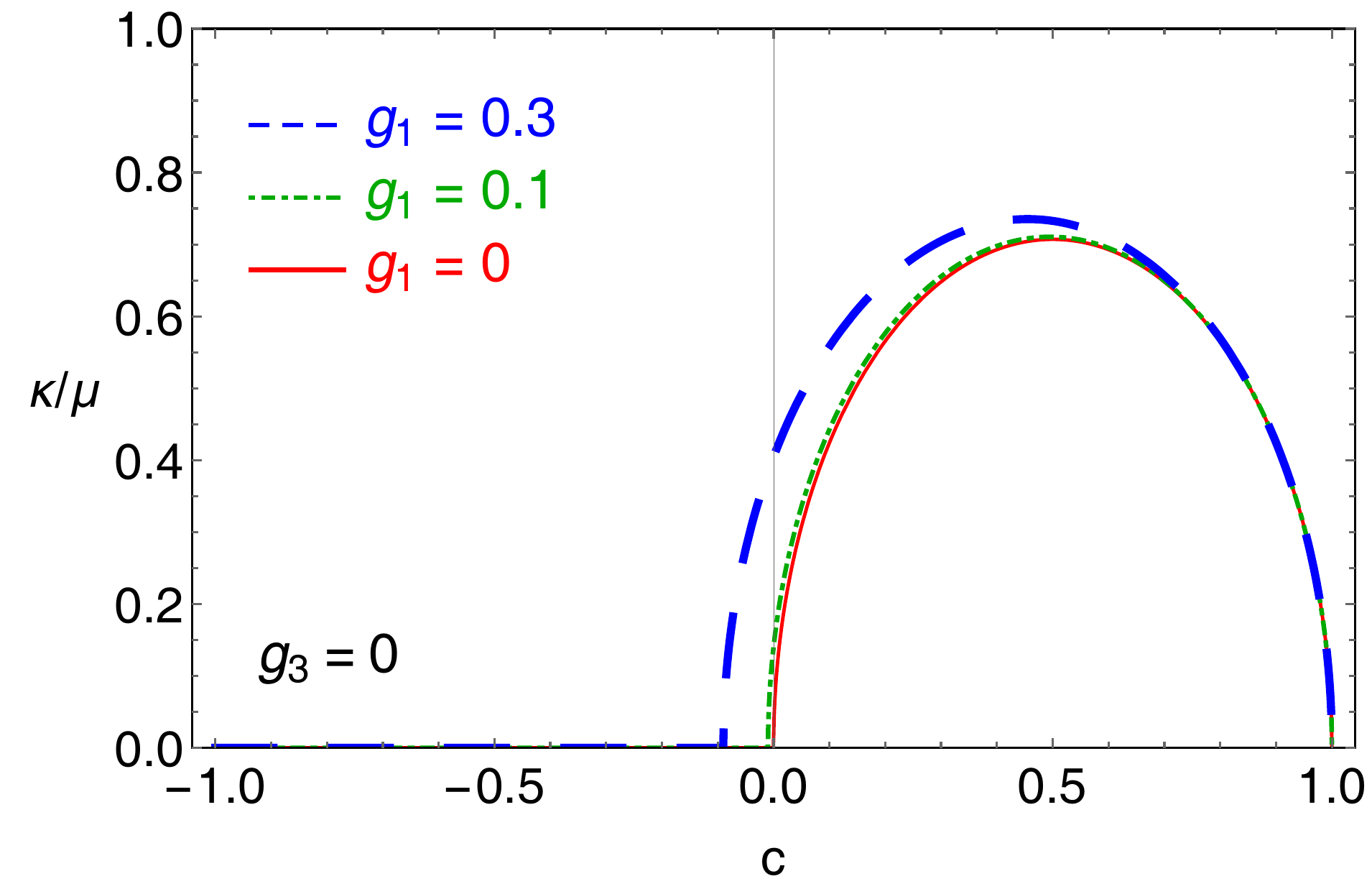}
\caption{Growth rates at $a=0$, as functions of $c\equiv\cos\theta$, for the left-right symmetry-breaking solution $(Q_-,\bar{Q}_-)$. Top: Only FP-NSSI, Bottom: Only FV-NSSI. }
\label{fig3}
\end{figure}
 
In the case of left-right symmetry breaking solution $(Q_-,\bar{Q}_-)$, non-zero complex eigenvalues are obtained for $\w/\mu\rightarrow 0$, indicating the presence of fast oscillations.
In Fig.\,\ref{fig2}, we show the growth rates in units of $\mu$ in the $a-\cos\theta$ plane, in the absence of NSSI (left panel), as well as in the presence of either FP-NSSI (centre panel) or FV-NSSI (right panel). The following observations can be made.

(i) In the absence of NSSI, one finds non-zero growth rates only for $\cos\theta>0$. This can be easily understood from Eq.\,(A1) and Eq.\,(A2) for $g_3=0$ and $g_1=0$ respectively. In both cases, the argument of the square-root is never negative for $\cos\theta\leq0$ and hence no instability occurs. In the scenario with instability, the growth rates do not depend on the sign of the neutrino-antineutrino asymmetry since the argument of the square-root depends on $a^2$ \cite{Chakraborty:2016lct}. 

(ii) The presence of FP-NSSI suppresses fast oscillations, shifting the non-zero growth rates to higher values of $\cos\theta$. Larger values of FP-NSSI shift the domain of fast oscillations to more acute-angle modes. This results in the effective pinching of the allowed region in the $a-\cos\theta$ parameter space.

(iii) The presence of FV-NSSI, on the other hand, expands the domain of fast oscillations. As can be seen from the right panel of Fig.\,\ref{fig2},
FV-NSSI can lead to fast oscillations even with negative $\cos\theta$, i.e., the $\nu$ and $\bar{\nu}$ modes with obtuse intersection angles start showing fast oscillations. This effect becomes significant for $g_1\sim\mathcal{O}(0.1)$.

In Fig.\,\ref{fig3}, we show the variation of the growth rates as a function of $\cos\theta$ for different values of $g_3$ 
(top panel) and $g_1$ (bottom panel). This gives a quantitative idea of the suppression and enhancement of the growth rate, with increasing $g_3$ and $g_1$, respectively.  Although both the plots are for zero neutrino-antineutrino asymmetry, we have checked that the features of the plot would remain unchanged with a non-zero asymmetry.

\begin{figure}[!t]
\includegraphics[scale=0.35]{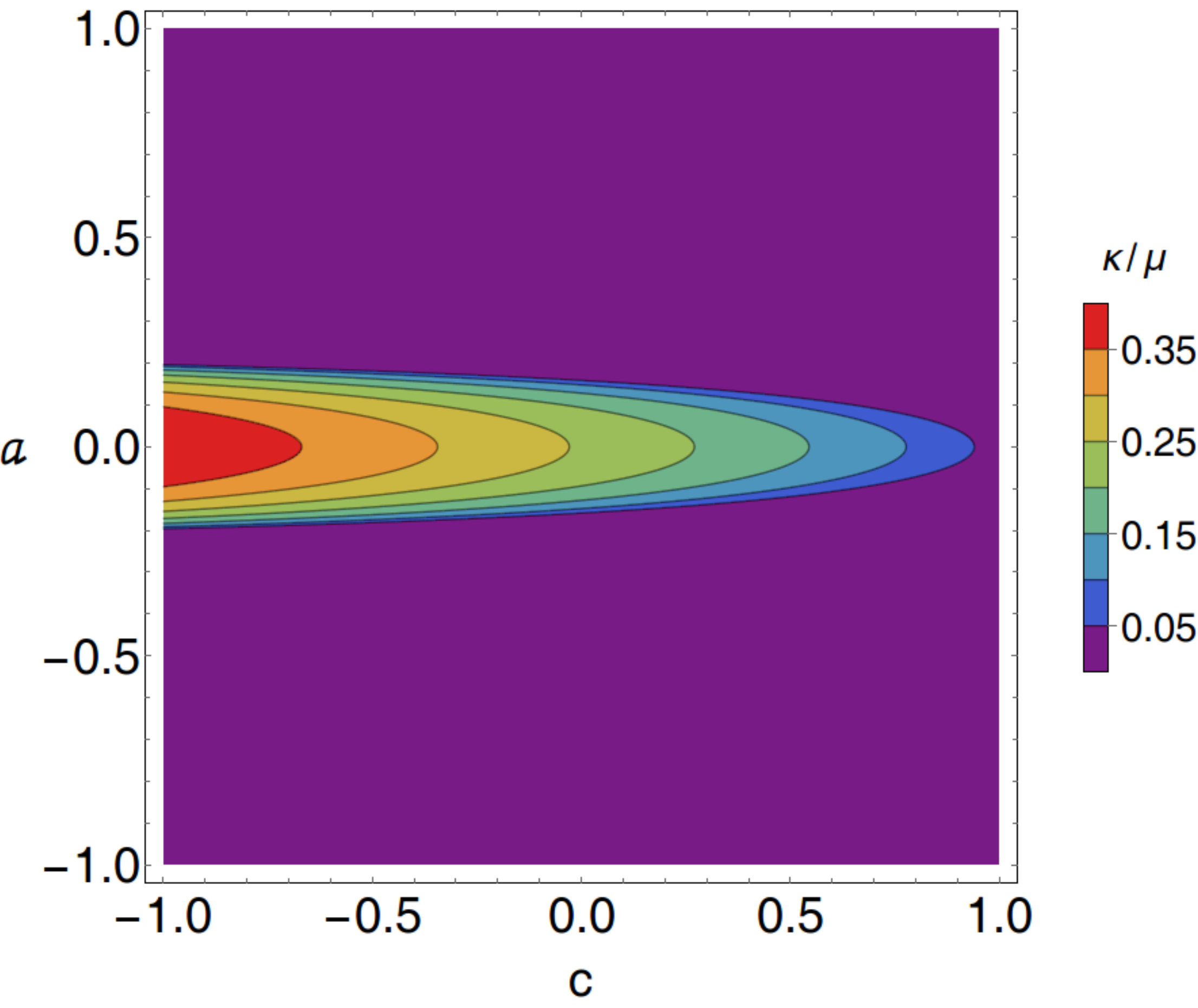}
\caption{Dependence of growth rates on $c\equiv\cos\theta$ and the $\nu-\bar{\nu}$ asymmetry $a$, for the left-right symmetric solution $(Q_+,\bar{Q}_+)$, for $g_1=0.1$ and $g_3=0$. }
\label{fig3a}
\end{figure}

\begin{figure}[!h]
\includegraphics[scale=0.4]{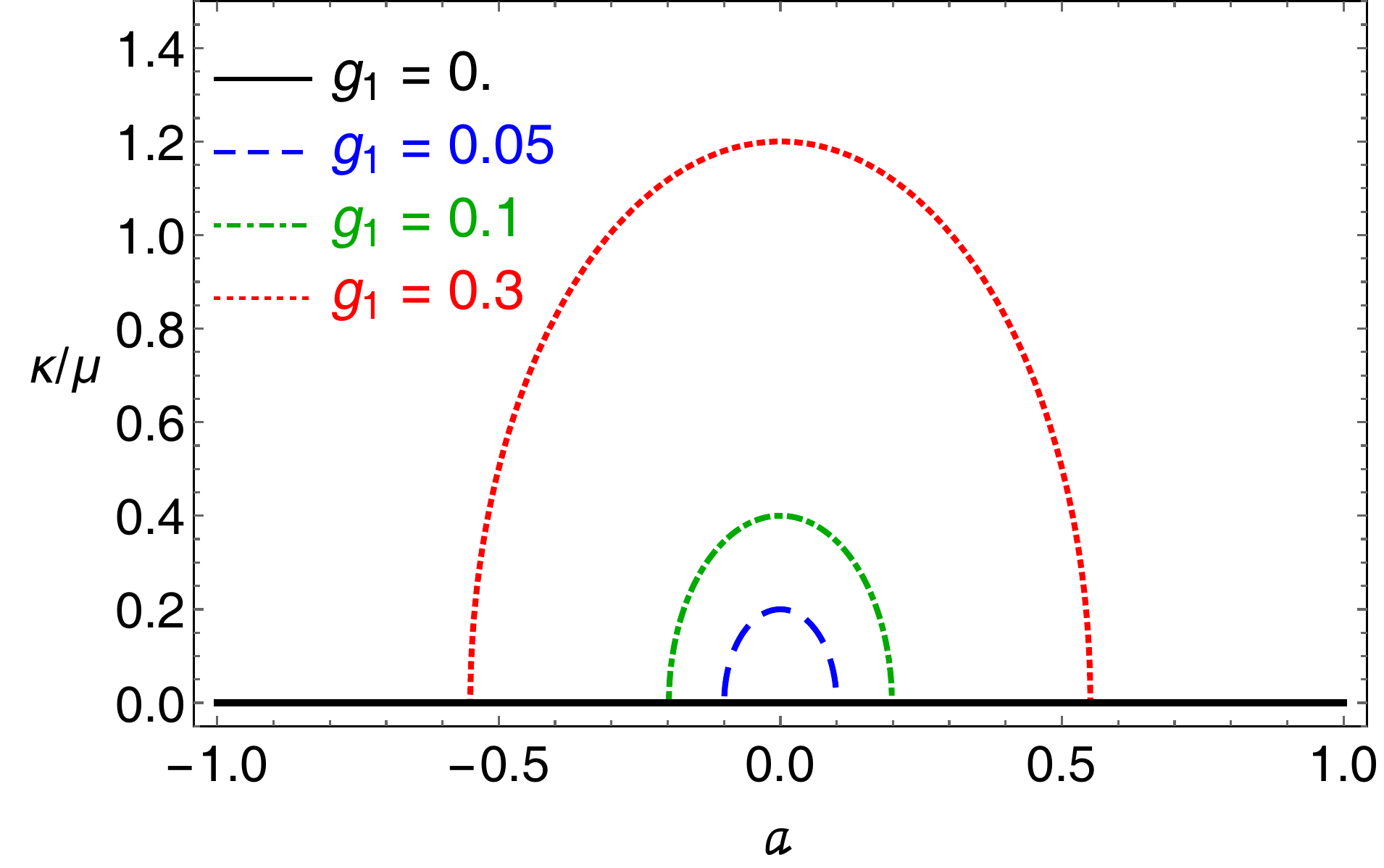}
\caption{ Growth rates in the effective two-beam scenario [the left-right symmetric solution $(Q_+,\bar{Q}_+)$ with $c=-1$], as functions of neutrino-antineutrino asymmetry $a$.}
\label{fig3b}
\end{figure}

In the case of the left-right symmetric solution $(Q_+,\bar{Q}_+)$, the SM predicts no fast oscillations, since the eigenvalues in the limit $\w/\mu\rightarrow 0$ are (see Appendix~\ref{app})
\begin{equation}\label{Qplg3}
\Omega^{\pm}_{\rm SM}=\dfrac{\mu}{2}\bigl[a(3-c)\pm\sqrt{a^2 (3-c)^2}\,\bigr]\,,
 \end{equation}
 which are always real. Also, in the presence of only FP-NSSI, the eigenvalues in Eq.\,(A3) can become complex only for $g_3\gtrsim\mathcal{O}(1)$.

 However, even with values of $g_1$ as small as $0.01$, it is possible to get complex eigenvalues for Eq.\,(A4).
 In Fig.\,\ref{fig3a}, we show the variation of the growth rates in the $a-\cos\theta$ plane for $g_1=0.1$. Clearly, large growth rates of $\mathcal{O}(\mu)$ are observed in the low-asymmetry region. 
 
 This opens up a new possibility for the back-to-back two-beam model, considered in \cite{Chakraborty:2016lct}, where spatial inhomogeneities were needed in order to start fast oscillations. In our intersecting four-beam model, the scenario corresponds to the left-right symmetric solution $(Q_+,\bar{Q}_+)$ with $c=-1$. In the presence of FV-NSSI, the eigenvalues are [see Eq.\,(A4)] 
\begin{equation}
 \Omega^{+}_{g_1}=2\mu\biggl[a\,\left(1- g_1^2\right)\pm \sqrt{ a^2 \left(1+g_1^2\right)^2-4 g_1^2}\biggr]\,.
\end{equation}
Though $g_1=0$ would exhibit no instabilities, even for small values of $g_1$, an instability would be developed for 
\begin{equation}
 |a|<\frac{2 g_1}{(1+g_1^2)}\,.
\end{equation}
Thus, no spatial inhomogeneities would be needed for fast oscillations as long as the neutrino-antineutrino asymmetry $a$ is sufficiently small.
In Fig.\,\ref{fig3b}, we show the growth rates as functions of $a$ for different values of $g_1$. Clearly larger $g_1$ values allow larger growth rates and  lead to instabilities even for larger asymmetries.

\subsection{Interplay of fast and slow oscillations and NSSI}
\label{sec:Non-linear}

\begin{figure}[!t]
\includegraphics[width=8cm,height=5cm]{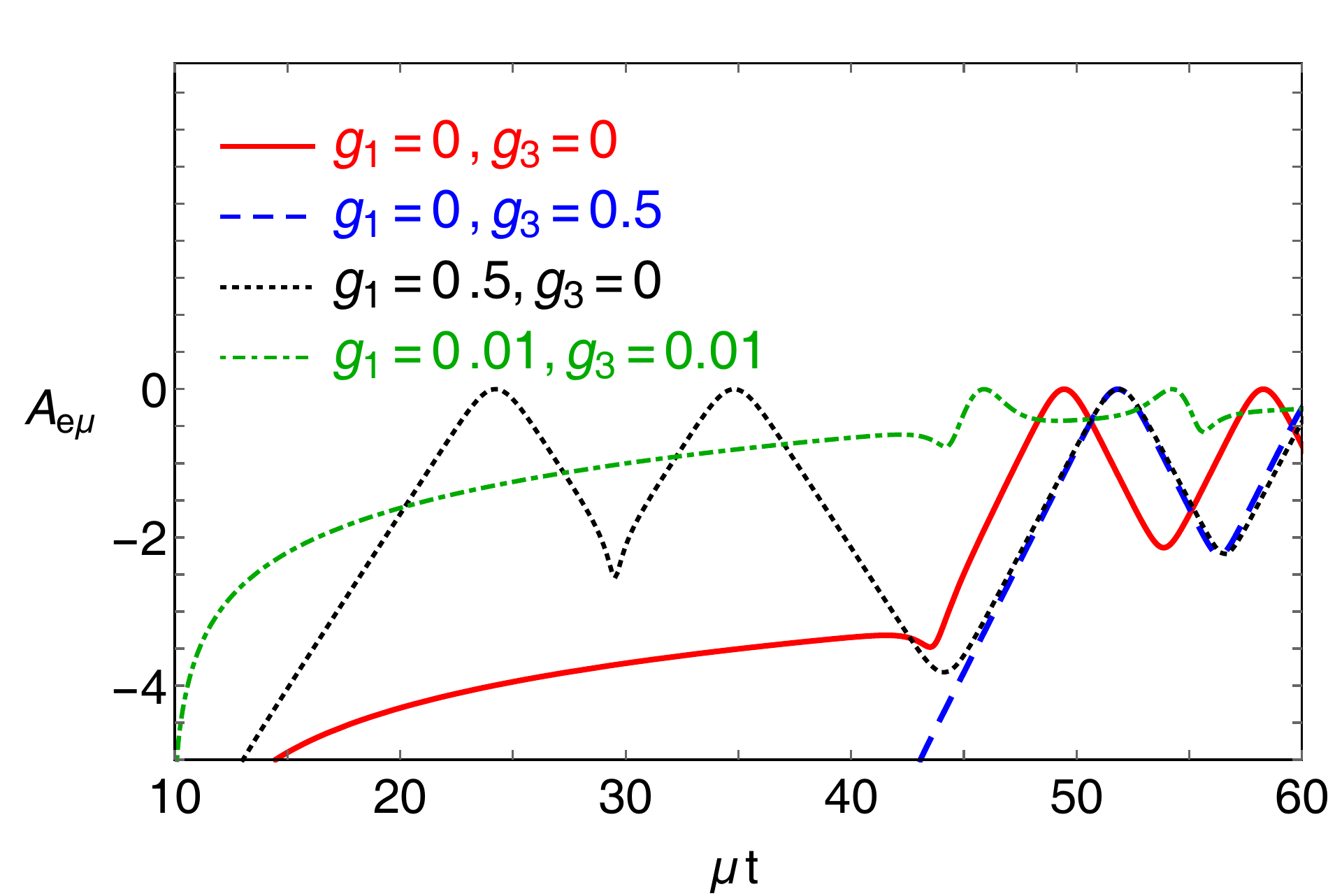}\\ \includegraphics[width=8.2cm,height=5.1cm]{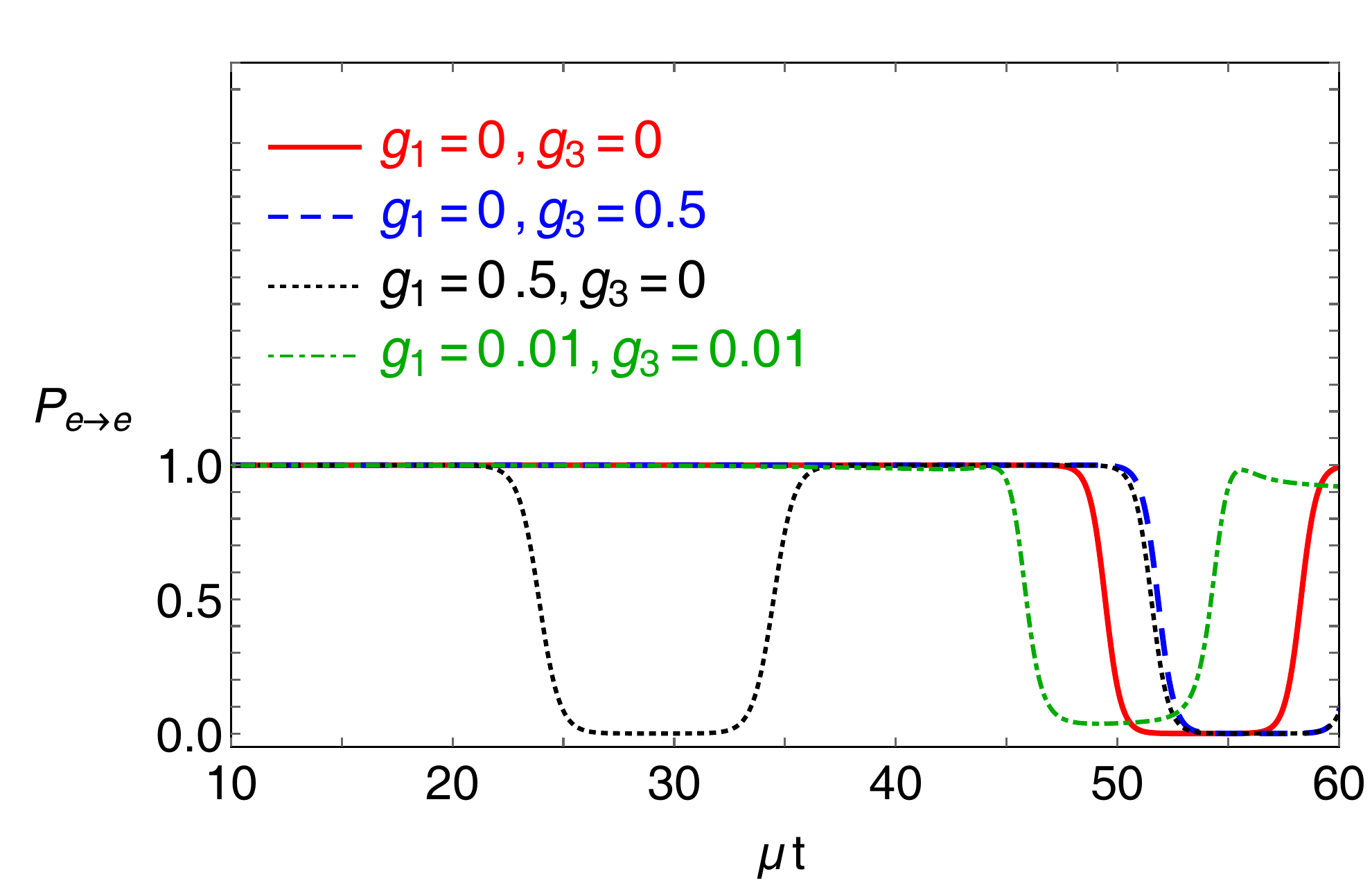}
\caption{Onset of fast oscillations from the numerical solution of the fully 
nonlinear EoM, for $a=0$ and $c=0.5$. The other parameters are chosen to be $\w/\mu_{\text{\tiny R}}=10^{-5}$, and $\vartheta_0=10^{-2}$. The top panel shows quantity $A_{e\mu}={\rm log}_{10}|S|$ which gives the extent of flavor conversion. The bottom panel shows the $\nu_e$ survival probabilities $P_{e\rightarrow e}$.}
\label{fig4}
\end{figure}

In this section, we demonstrate the effect of NSSI 
by numerically solving the fully non-linear equations of motion for the intersecting four-neutrino system. The results relevant for the onset of fast oscillations are shown in Fig.\,\ref{fig4}. 
Note that the magnitude of the off-diagonal parameter $|S|$ is same for all the modes, and characterizes flavor conversions. 
The top panel shows the quantity $A_{e\mu}\equiv{\rm log}_{10}|S|$, which gives the amplitude of the flavor conversions, while the bottom panel shows $P_{e\rightarrow e}$, the $\nu_e$ survival probability. The following observations may be made.

(i) The time evolution of $A_{e\mu}$ in the SM shows an initial flat phase, followed by a sharp rise. This sharp rise corresponds to the onset of fast oscillations. The initial flat phase is an effect of non-zero $\w$. It does not succeed in causing large flavor conversions as can be seen from $P_{e\rightarrow e}$ in the bottom panel of Fig.\,\ref{fig4}. 

(ii) As expected from the linear analysis, a non-zero $g_3$ delays the onset, whereas a non-zero $g_1$ reduces this initial waiting period. These effects start becoming appreciable when $g_1\,,g_3\gtrsim\mathcal{O}(0.1)$.

(iii) The growth rate obtained from the numerical simulation is in good agreement with the calculations from the linear stability analysis.

(iv) An eigenvalue equation was not possible when both FP-NSSI and FV-NSSI are present. However, the numerical solution verifies that the initial part of the dynamics is governed by a linear growth, as expected. As can be seen from $P_{e\rightarrow e}$, significant flavor conversion starts happening only when the  exponential growth of oscillations takes over.

(v) Note that fast flavor oscillations are only sensitive to the asymmetries in the angular distributions of the neutrino-antineutrino beams, and energy only affects the onset, and not the rate. The subleading terms of order $\omega/\mu$ act as a seed necessary for the onset of the oscillations. This does not affect the rate of oscillations, but only changes the initial waiting period \cite{Dasgupta:2017oko}.

\begin{figure}[!t]
\includegraphics[width=8cm,height=5cm]{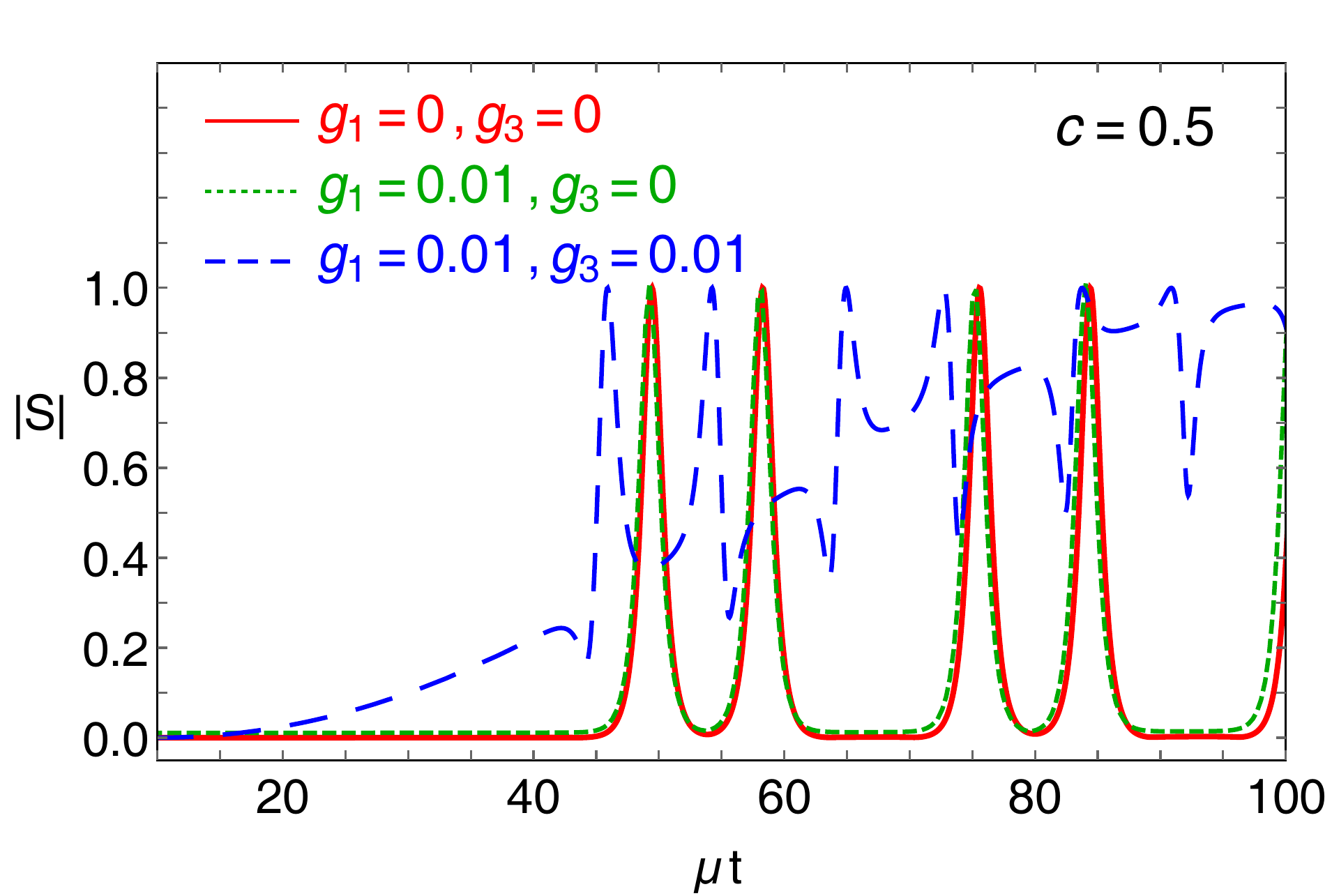}\\ \includegraphics[width=8cm,height=5cm]{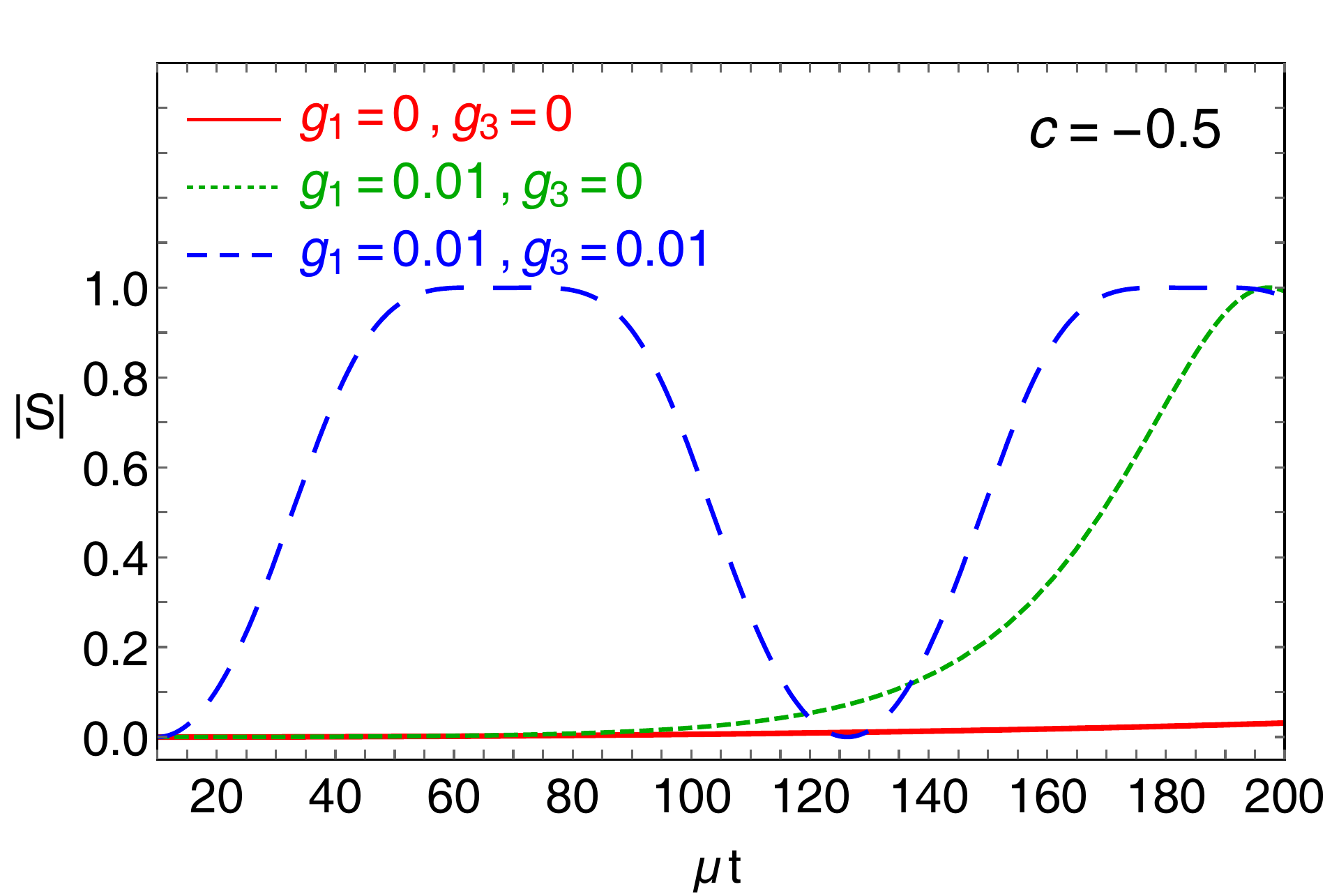}
\caption{Onset of oscillation in terms of the off-diagonal parameter $|S|$ from the numerical solution of the fully
nonlinear EoM for $a=0$. The top panel shows the plot for $c=0.5$ where we expect fast oscillations. The bottom panel shows the case $c=-0.5$, when fast oscillations are absent. The other parameters are chosen to be $\w/\mu_{\text{\tiny R}}=10^{-5}$, and $\vartheta_0=10^{-2}$.}
\label{fig5}
\end{figure}

\begin{figure*}
\includegraphics[scale=0.45]{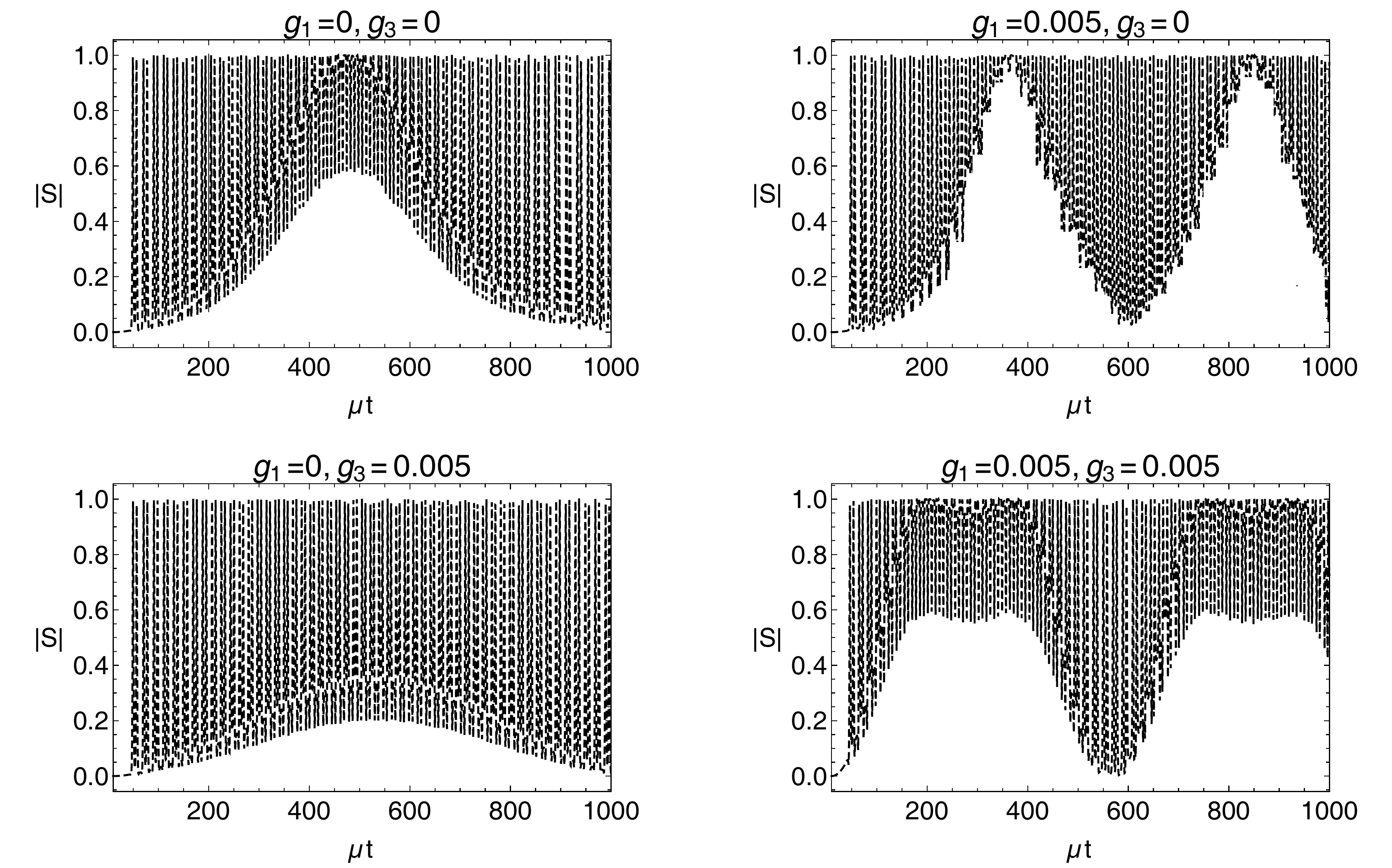}
\caption{Long-time behavior of collective oscillations in terms of the off-diagonal parameter $|S|$ from the numerical solution of the fully
nonlinear EoM for $a=0$ and $c=0.5$. The other parameters are chosen to be $\w/\mu_{\text{\tiny R}}=10^{-4}$, and $\vartheta_0=10^{-2}$. The rapid variations correspond to the fast oscillation frequency while the slowly changing envelope correspond to the slow oscillation frequency. The comparison of two columns indicates the effect of $g_1$ while the comparison of the rows indicates the effects of $g_3$. }
\label{fig6}
\end{figure*}

This point is further elucidated in the top panel of Fig.\,\ref{fig5}, where we plot the quantity $|S|$. 
Small values of $g_1$, in the absence of $g_3$, do not affect the fast oscillations at all. On the other hand, when both $g_1$ and $g_3$ are non-zero, even when their values are $\mathcal{O}(0.01)$, an early linear rise as well as an early onset of exponential growth of oscillations may be observed.
The fast oscillations seen at slightly later times seem to be riding on a slowly rising curve. This turns out to be a combined effect of slow oscillations and the early linear rise.

The bottom panel of Fig.\,\ref{fig5} demonstrates the effect of NSSI on slow oscillations when fast oscillations are absent. Clearly, $g_1\sim\mathcal{O}(0.01)$, even in the absence of $g_3$, can shift the onset of slow oscillations to much earlier times. With the addition of a similar magnitude of $g_3$, the onset may be further hastened, bringing it in the domain of the onset of fast oscillations. Thus, it is possible for the oscillations to start deep inside the core, still keeping their frequencies small.

The long-time behavior of the non-linear oscillations would be an interplay of the fast and slow oscillations, influenced by NSSI. In Fig.\,\ref{fig6}, we show this long-time evolution of $|S|$ for different cases.

(i) In the SM $(g_1=0\,,g_3=0)$, one can see fast oscillations with frequency $\sim\mu=10^4\,{\rm km}^{-1}= 0.03\,{\rm s}^{-1}$, modulated by an envelope of slow oscillations with frequency $\sim\sqrt{\w\mu}\approx100\,{\rm km}^{-1}=0.0003\,{\rm s}^{-1}$. 

(ii) Small values of $g_1\sim0.005$ affect the slow oscillation frequency appreciably, while keeping the fast oscillations relatively unchanged. The modulating envelope increases in magnitude as well as frequency if $g_1$ is increased.

(iii) Non-zero values of $g_3$ does not affect the frequencies of fast and slow oscillation. However the modulating envelope 
decreases in amplitude. This is true both in the presence and absence of $g_1$.

Larger values of NSSI also start affecting the frequency of fast oscillations. However, we will not explore such scenarios in this paper.

\section{Summary and Conclusions}
\label{sec:Conclusions}

In this paper, we have investigated the effects of NSSI of neutrinos on collective flavor oscillations of neutrinos exiting a SN. We employ a linearized stability analysis, using a small-amplitude expansion in the off-diagonal element $S_{\w,u}$ of the density matrix in the flavor basis. In particular, we explore how the exponentially growing flavor conversion modes are affected by the NSSI. We characterize the NSSI with the help of two parameters, the flavor-preserving (FP-NSSI) parameter $g_3$ and the flavor-violating (FV-NSSI) parameter $g_1$.

While the linear stability analysis in SM leads to an eigenvalue equation, the most general EoM incorporating NSSI does not do so. Therefore, the problem of onset of oscillations, in complete generality, cannot be solved analytically. However, we find that, if the system has only FP-NSSI, we get a straightforward eigenvalue equation, whose complex eigenvalue directly signals an instability. 
In order to illustrate the NSSI effects on bipolar collective oscillations, we present the analysis for a two-box spectrum. We analytically demonstrate the suppression of collective oscillations by FP-NSSI, explaining the results from previous literature. If only FV-NSSI are present, an approximate eigenvalue equation can still be obtained, which qualitatively motivates the enhanced growth rates observed in numerical simulations. In the presence of both kinds of NSSI, one gets an additional linear contribution to $S_{\w,u}$, which further affects its evolution.

In order to illustrate the NSSI effects on fast flavor oscillations, we focus on the intersecting four-beam model, which is the simplest model where such effects can be analytically studied. Here two beams each of neutrinos and antineutrinos, with a neutrino-antineutrino asymmetry $a$, intersect at an angle $\theta$. Using the linear stability analysis, we study the left-right symmetric and the left-right symmetry-breaking modes of this system.
For the left-right symmetry-breaking modes, it is observed that the instability is restricted to a smaller region in the $a-\cos\theta$ plane with FP-NSSI, while FV-NSSI widens the corresponding region. Moreover, while SM allows an instability only for $\cos\theta>0$, the presence of FV-NSSI allows an instability even for an obtuse angle $\theta$.

A striking corollary of the last result is that fast oscillations can take place for the two-beam system consisting of opposing neutrino and antineutrino beams, even in the absence of inhomogeneities. Indeed this scenario is equivalent to the left-right symmetric solution of the intersecting four-beam model with $\cos\theta=-1$, where instability can be developed in the presence of FV-NSSI. This is in stark contrast to previous results in SM, where the two-beam system could exhibit fast oscillations only if spatial inhomogeneities were present. Clearly, the lepton flavor universality-breaking NSSI couplings now play the role of the symmetry-breaking seed required to give rise to an instability.

We also solve the complete non-linear EoMs numerically for the four-beam system, for the onset of oscillations as well as long-time behavior. It is observed that when both FP-NSSI and FV-NSSI are present, the extra linear contribution to $S$ results in an initial linear growth, which may later be dominated by fast oscillations. However, in the situations where fast oscillations are absent, the same linear term helps in bringing the onset of slow collective oscillations to significantly earlier times. 

The long-time behavior of the system may be described by fast collective oscillations modulated by the slow ones. It is observed that FP-NSSI suppress the amplitude of these modulations while FV-NSSI enhance their amplitude and frequency.

Since NSSI can bring the fast as well as slow oscillations nearer to the core, they can have important consequences for the explosion mechanism and nucleosynthesis as flavor conversions can start occurring earlier. This indicates the importance of going beyond the approximation of a neutrinosphere and taking care of collisions in a neutrino oscillation analysis. Note that the effect of NSSI on the neutronization burst as calculated in \cite{Das:2017iuj} still stays valid.

This work uses a two-flavor framework and demonstrates fast oscillations using the intersecting four-beam model. A more complete analysis would require a generalization to a three-flavor framework and continuous spectra. It would also allow us to address some important questions of principle.
In a previous work \cite{Das:2017iuj}, we had shown that the requirement for a crossing in the spectrum for the development of collective oscillations is not necessary in the presence of FV-NSSI. It might be interesting to check whether the requirement of a crossing in the angular spectra for the onset of fast oscillations would survive even in the presence of FV-NSSI.

\section*{Acknowledgements} 
We would like to thank Basudeb Dasgupta for insightful discussions, suggestions, and comments on the manuscript. 
The work of M.S. was supported by the Max-Planck Partnergroup ``Astroparticle Physics" of the Max-Planck-Gesellschaft awarded to Basudeb Dasgupta. This project has received partial support from the European Union's Horizon 2020 research and innovation programme under the Marie-Sklodowska-Curie grant agreement Nos. 674896 and 690575.

\onecolumngrid
\appendix

\section{Glossary of symbols}
\label{glossary}

Table \ref{tab2} gives a list of the some of the selected symbols used in this paper, with the descriptions of their meanings, for clarity.

\begin{table}[!h]
\centering
  \begin{tabular}{|p{2.4cm}|p{15cm}|}
        \hline
      ~~Symbol~                                            &~~~~~~~~~~~~~~~~~~~~~~~~~~~~~~~~~~~~~~~~~~~~~~~~~~~Description\\
	\hline                     
           ~$G_F$                &~ The Fermi constant\\
           ~$G$                  &~ The Coupling matrix in the flavor basis, in the presence of NSSI \\
	  ~$\gamma^\mu$          &~ The Dirac Matrix\\
	  ~$\gamma_{ee},\gamma_{ex},\gamma_{xx}$         &~ The NSSI components of the coupling matrix in the flavor basis  \\
	  ~$\gamma$              &~ The real part of $\Omega$, where $S=Q\,e^{-i\,\Omega t}$.\\
        ~$g_0,\,g_1,\,g_2,\,g_3$            &~ The NSSI couplings such that $G=\frac{1}{2}\left(g_0{\mathbb I}+g_1\sigma_1+g_2\sigma_2+g_3\sigma_3\right)$\\
	 ~${\boldsymbol g}$             &~ The 3-vector of NSSI couplings: ${\boldsymbol g}\equiv(g_1,\,g_2,\,g_3)$\\
	 ~$g_{\w,u}$            &~ The difference of neutrino flavor spectra as a function of $\w$ and $u$ [Eq.\,(\ref{Diffspectrum})]\\
	 ~$\mu_{\text{\tiny R}}$ &~ The value of $\mu$ at the neutrinosphere\\

	 ~$\widetilde{\mu}_r$    &~ $\widetilde{\mu}_r \equiv \mu_{\text{\tiny R}}\,R^4/(2\,r^4)\,$\\	
	~$\lam_r$                &~  The matter potential at a radius $r$: $\lam_r \equiv \sqrt{2}\,G_F\,n_e(r)$\\
	~$\widetilde{\lam}_r$     &~ $\widetilde{\lam}_r\equiv\lam_r\,R^2/(2\,r^2)$\\
	~$\overline{\lam}_r$      &~ The effective matter term $\overline{\lam}_r\equiv\lam_r+u\bigl[\widetilde{\lam}_r+\widetilde{\mu}_r\,\epsilon\,(1-g_1^2+3g_3^2+4g_3)\bigr]$\\
	                          
	\hline
  \end{tabular}
  \caption{List of some of the selected symbols used in this article, and their meanings.}
   \label{tab2}
 \end{table}

\section{ The eigenvalues of the Hamiltonian in the four-beam model}
\label{app}

The expressions of the $\mathcal{H}$ matrix elements in Eq.\,(\ref{eigeqn}), for the L-R symmetric and symmetry breaking solutions, are given below. 

\subsection{L-R Symmetric solution $(Q_+,\bar{Q}_+)$}
For $g_1=0\,,g_3\neq0$, one has
\begin{eqnarray}
 \mathcal{H}_{11}&=&\w+\frac{\mu}{2}\biggl[(1+g_3)\biggl(c-3+g_3(-5+7c)+a\left(3-c+13g_3+cg_3\right)\biggr)\biggr]\,,\nonumber\\
 \mathcal{H}_{12}&=&\frac{\mu}{2}(1- a) (3-c) (1 - g_3^2)\,,\nonumber\\
 \mathcal{H}_{21}&=&-\frac{\mu}{2}(1 + a) (3-c) (1 - g_3^2)\,,\nonumber\\
 \mathcal{H}_{22}&=&-\w+\frac{\mu}{2}\biggl[(1+g_3)\biggl((1+a)(3-c)+g_3\left(5-7c+13 a+a c\right)\biggr)\biggr]\,.\nonumber
 \end{eqnarray}

Similarly, for $g_1\neq0\,,g_3=0$, one has 

\begin{eqnarray}
 \mathcal{H}_{11}&=&\w+\frac{\mu}{2}\,\biggl[g_1^2-3 c g_1^2+c-3-a \left((c+5) g_1^2+c-3\right)\biggr]\,,\nonumber\\
 \mathcal{H}_{12}&=&\frac{\mu}{2}(1 - a) (3-c) (1 + g_1^2)\,,\nonumber\\
 \mathcal{H}_{21}&=&-\frac{\mu}{2}(1 + a) (3-c) (1 + g_1^2)\,,\nonumber\\
 \mathcal{H}_{22}&=&-\w-\frac{\mu}{2}\biggl[(1+a) (c-3)+g_1^2 (1-3c+a\,(c+5))\biggr]\,.\nonumber
 \end{eqnarray}

\subsection{L-R Symmetry breaking solution $(Q_-,\bar{Q}_-)$}
For $g_1=0\,,g_3\neq0$, one has
\begin{eqnarray}
 \mathcal{H}_{11}&=&\w+\mu\biggl[\frac{1}{2}(1+a)(1+c)\left(1-g_3^2\right)+(1+4 g_3+3 g_3^2)(2 a+c-1)\biggr]\,,\nonumber\\
 \mathcal{H}_{12}&=&-\frac{\mu}{2}(1 - a) (1 + c) (1 - g_3^2)\,,\nonumber\\
 \mathcal{H}_{21}&=&\frac{\mu}{2}(1 + a) (1 + c) (1 - g_3^2)\,,\nonumber\\
 \mathcal{H}_{22}&=&-\w+\mu\biggl[-\frac{1}{2}(1-a)(1+c)\left(1-g_3^2\right)+(1+4 g_3+3 g_3^2) (2 a-c+1)\biggr]\,.\nonumber
 \end{eqnarray}

Similarly, for $g_1\neq0\,,g_3=0$, one has 

\begin{eqnarray}
 \mathcal{H}_{11}&=&\w+\mu\biggl[\frac{1}{2}(1+a)(1+c)\left(1+g_1^2\right)+(1-g_1^2)(2 a+c-1)\biggr]\,,\nonumber\\
 \mathcal{H}_{12}&=&-\frac{\mu}{2}(1 - a) (1 + c) (1 + g_1^2)\,,\nonumber\\
 \mathcal{H}_{21}&=&\frac{\mu}{2}(1 + a) (1 + c) (1 + g_1^2)\,,\nonumber\\
 \mathcal{H}_{22}&=&-\w+\mu\biggl[-\frac{1}{2}(1-a)(1+c)\left(1+g_1^2\right)+(1-g_1^2) (2 a-c+1)\biggr]\,.\nonumber
 \end{eqnarray}

The corresponding eigenvalues, in the limit $\w/\mu\rightarrow0$, are listed in Table \ref{tab1}. We put $\w=0$ since we are only interested in the coefficient of $\mu$ for fast oscillations.

\begin{table}
\centering
  \begin{tabular}{|p{2.5cm}|p{15cm}|}
         \hline
      ~~~~~Solution~~                                            &~~~~~~~~~~~~~~~~~~~~~~~~~~~~~~~~~~~~~~~~~~~~~~~~~~~Eigenvalue\\
           \hline                                                &~\\
	  ~$(Q_-,\bar{Q}_-)$                          &$~\Omega^{-}_{g_3}=\dfrac{\mu}{2}\biggl\{\,a\bigl[5+c+16g_3+(11-c)g_3^2\bigr]\pm\sqrt{a^2(1+c)^2(1-g_3^2)^2-8(1-c)(1+g_3)^2(1+3g_3)\bigl[c(1+g_3)-2g_3\bigr]}\,\biggr\}$\\
           $~g_1=0\,,g_3\neq0$                        & ~~~~~~~~~~~~~~~~~~~~~~~~~~~~~~~~~~~~~~~~~~~~~~~~~~~~~~~~~~~~~~~~~~~~~~~~~~~~~~~~~~~~~~~~~~~~~~~~~~~~~~~~~~~~~~~~~~~~~~~~~~~~~~~~~~~~~(A1)\\
         \hline                                          	 &~~\\     
       ~$(Q_-,\bar{Q}_-)$                             &$~\Omega^{-}_{g_1} =\dfrac{\mu}{2}\biggl\{\,a\bigl[5+c-g_1^2(3-c)\bigr] \pm\sqrt{a^2(1+c)^2(1+g_1^2)^2-8(1-c)(1-g_1^2)(c+g_1^2)}\,\biggr\}$~~~~~~~~~~~~~~~~~~~~~~~(A2)\\
        $~g_1\neq0\,,g_3=0$                                  &\\
         \hline                                                  &~~\\
  ~$(Q_+,\bar{Q}_+)$                                          &$~\Omega^{+}_{g_3}=\dfrac{\mu}{2}(1+g_3)\biggl\{a\bigl[3- c(1-g_3)+13 g_3\bigr]\pm\sqrt{a^2 (3-c)^2 (1-g_3)^2+16 g_3(1-c)\bigl[3-c+g_3(1-3c)\bigr]}\biggr\}$~(A3)\\
     $~g_1=0\,,g_3\neq0$                                   &~~\\
        \hline                                                   &\\
       ~ $(Q_+,\bar{Q}_+)$                                   &$~ \Omega^{+}_{g_1} =\dfrac{\mu}{2}\biggl\{-a \bigl[(c+5)g_1^2+c-3\bigr]\pm\sqrt{a^2(c-3)^2(1+g_1^2)^2-8g_1^2(1-c)\bigl[(1+c)g_1^2-c+3\bigr]}\biggr\} $~~~~~~~~~~~~~~~(A4)\\
        $~g_1\neq0\,,g_3=0$                                    &\\

         \hline
  \end{tabular}
  \caption{Eigenvalues of Eq.\,(\ref{eigeqn}) for the four cases described in this Appendix, with $\w=0$.}
   \label{tab1}
 \end{table}


\twocolumngrid

\end{document}